\documentclass[aps,prc,twocolumn,amsmath,superscriptaddress,showpacs]{revtex4-2}
\usepackage{graphicx}
\usepackage[usenames]{color}
\usepackage{isotope}
\usepackage{hyperref}
\usepackage{comment}
\usepackage{multirow}
\usepackage{float}

\newcommand{\ba}{\begin{eqnarray}}
\newcommand{\ea}{\end{eqnarray}}
\newcommand{\beq}{\begin{equation}}
\newcommand{\eeq}{\end{equation}}
\newcommand{\be}{\begin{equation}}
\newcommand{\ee}{\end{equation}}

\begin{document}
\title{Nuclear binding, correlations, and the $A$-dependence of the EMC effect}
\author{Omar Benhar}
\email{omar.benhar@roma1.infn.it}
\affiliation{INFN, Sezione di Roma, 00185 Roma, Italy}
\author{Alessandro Lovato}
\email{lovato@anl.gov}
\affiliation{Physics Division, Argonne National Laboratory, Argonne, Illinois 60439, USA} 
\affiliation{Computational Science Division, Argonne National Laboratory, Argonne, Illinois 60439, USA}
\affiliation{INFN-TIFPA Trento Institute for Fundamental Physics and Applications, 38123 Povo, Italy}
\affiliation{Instituto de Física Corpuscular (IFIC), Consejo Superior de Investigaciones Científicas (CSIC) and Universidad de Valencia
E-46980 Paterna, Valencia, Spain}
%
\begin{abstract}
The measurements of inclusive electron scattering from nuclear targets 
carried out at the Thomas Jefferson National Accelerator Facility in the mid 2000s have provided 
valuable novel information on the $A$-dependence of the modifications of nuclear 
structure functions known as EMC effect.
We argue that these data are best described in terms of the scaling variable 
$\widetilde{y}$, designed to take into account dynamical effects in interacting 
many-particle systems, and analyse the $A$-dependence of the slope of the inclusive cross section ratios, $R_A = (\sigma_A/A)/(\sigma_2/2)$, providing a measure of the size of the EMC effect in the region where nuclear binding plays a leading role. The results of our study clearly hint at a  linear correlation between $dR_A(\widetilde{y})/d\widetilde{y}$ and the average  nucleon removal energy $\langle E_A \rangle$. The role of correlation effects 
 in the determination of $\langle E_A \rangle$ is highlighted.  
\end{abstract}
%
\date{\today}
\maketitle
%
\paragraph*{Introduction} The EMC effect\textemdash named after the European Muon Collaboration, and first reported at the beginning of the  1980s~\cite{CERN_courier_1982,EMC_1983}\textemdash was the observation
that the electromagnetic structure functions per nucleon of iron and deuterium, measured at CERN using a 280~GeV muon beam, displayed a surprising pattern of differences, clearly visible as deviations from unity in their ratio.
More generally, the term EMC effect has later come to refer to any departure from unity of the 
ratio $R_A$ between the cross section per nucleon of a nucleus of mass number $A$ and 
that of the deuteron, corresponding to $A=2$. 

A large set of data collected at CERN~\cite{Bari85,Ashman88}, SLAC~\cite{Gomez94}, HERA~\cite{Ackerstaff2000,Airapetian2003} and Jefferson Lab~\cite{Seely09,Arrington12,CLAS:2019,Arrington21,Karki2023} unambiguously confirmed that for values of the Bjorken scaling variable in the range $0.35 < x < 0.70$ $R_A$ exhibits a nearly linear behaviour, and decreases from $\sim$1 to a minimum which, depending on $A$, can be as low as $\sim$0.8. Such a large effect was totally unexpected, because the binding energy of nucleons in atomic nuclei is small\textemdash in fact, altogether negligible\textemdash compared to the large energy transfer associated 
with Deep Inelastic Scattering (DIS) processes.

No definitive explanation has emerged, mainly because most existing theoretical approaches are 
inherently inadequate to provide a fully quantitative account of the data over the whole range of $x$. The degree of understanding of the EMC effect\textemdash or lack thereof\textemdash  has been brilliantly summarised by the authors of a paper published in the CERN Courier in 2013, entitled 
{\it The EMC effect still puzzles after 30 years}~\cite{CERN_courier_2013}.

Many different models, extensively reviewed in, e.g., Refs.~\cite{Arneodo94,Geesaman95,Norton03,Malace:2014}, have been proposed to explain the reduction of $R_A$ at intermediate $x$. The role of nuclear binding has been often, although not always, taken into account, but the results of these studies generally suggest that the effect of binding alone is not large enough to describe the data; see, e.g., Refs.~\cite{Ciofi91,Benhar97}. 

In 2009, the interest in nuclear effects in DIS was reignited by the appearance of new data taken at Jefferson Lab\textemdash hereafter JLab\textemdash by the E03-103 Collaboration~\cite{Seely09}. These authors carried out a detailed analysis of the $A$-dependence of the EMC effect using inclusive cross sections measured by scattering a 5.766 GeV electron beam off \isotope[2][]{H}, \isotope[3][]{He}, \isotope[4][]{He}, \isotope[9][]{Be}, and \isotope[12][]{C}.
In order to minimise normalisation uncertainties, the size of the effect was conveniently characterised by the slope of the cross section ratio, $dR_A/dx$, in the linear region. The most striking outcome of this study was the observation of a significant departure of the data point corresponding to \isotope[9][]{Be} from the linear dependence on the average nuclear density emerging from the fit of 
SLAC data of Ref.~\cite{Gomez94}. According to Seely~{\it et~al.}~\cite{Seely09}, this anomaly may be explained considering that, because the nucleus of \isotope[9][]{Be} consists of two tightly bound alpha particles and an additional neutron, the nucleon interacting with the beam particle is more likely found in a region of density significantly higher than the {\it average} beryllium density. 

The data of Ref.~\cite{Seely09}, suggesting that the size of the EMC effect is driven by the {\it local}, rather than the {\it average}, nuclear density, imply in turn that short-range dynamics, responsible for the appearance of high-momentum nucleons which largely determine the nuclear structure functions at $x>1$, may as well play a significant role at intermediate $x$~\cite{Weinstein11,Hen12}. 
Based on this conjecture, Fomin~{\it et~al.}~\cite{Nadia2012} investigated the dependence 
of the slopes of the EMC ratios on a parameter dubbed $R_{2N}$, designed to 
provide a measure of the ratio between the probabilities of finding a nucleon belonging to a high momentum pair in a nucleus of mass number $A$ and in the deuteron. A somewhat similar study, 
aimed at determining the dependence of the slope on the average number of nucleons involved in local 
fluctuations of the nuclear density, measured by a parameter referred to as $a_2$, has been performed by 
the authors of Ref.~\cite{Arrington12}. While supporting the local density interpretation proposed by Seely {\it et al.}~\cite{Seely09}, 
the results of this study do not rule out the alternative explanation of Refs.~\cite{Weinstein11,Hen12,Nadia2012}, advocating the role of
high-momentum nucleon pairs. 
The results of  a more recent analysis carried out by Arrington and Fomin~\cite{Arrington2019} also led to the conclusion
that  the connection between the EMC effect and short-range correlations (SRC)\textemdash with the ensuing appearance of a significant  isospin dependence~\cite{Fomin:2026swt}\textemdash is still lacking a firm understanding. 

Arrington {\it et al.}~\cite{Arrington12}, also investigated a somewhat different explanation of the EMC effect. 
These authors analysed the dependence of the slopes of the cross section ratio on the average nucleon removal energy $\langle E_A \rangle$, 
and found a remarkable linear correlation. 
However, they did not pursue this observation further, nor did they discuss thoroughly the role of 
SRC in the determination of $\langle E_A \rangle$.

In this Letter\textemdash aimed at extending and improving the pioneering study 
of Ref.~\cite{Benhar2012}\textemdash the $A$-dependence of the EMC effect and the role of 
the nucleon removal energy are analysed using the scaling 
variable ${\widetilde y}$, originally proposed in Ref.~\cite{Benhar2000}. The distinctive feature of this approach is that, unlike the scaling variables routinely employed to describe DIS data, ${\widetilde y}$ has a clear physical interpretation in the target rest frame, in which nuclear dynamics can be accurately described, and allows for a straightforward identification of the effect of nuclear binding.~\\

\paragraph*{Emergence of $\widetilde y$-scaling}

Consider a process whereby the probe particle is scattered by a many-body system at rest. At momentum transfer ${\bf q}$ such that $|{\bf q}| \gg d^{-1}$, 
with $d$ being the average distance between target constituents, the impulse approximation regime sets in, and scattering off individual 
constituents\textemdash the mass and momentum of which will be denoted $m$  and ${\bf p}$, respectively\textemdash is the dominant reaction mechanism. 
The interaction with the probe triggers a transition of the target from the ground state to a 
final state $|F\rangle$ comprising  the struck constituent with momentum  ${\bf p}+{\bf q}$ and the spectator system, left in the state $|{\mathcal R} \rangle$ with energy $E_{\mathcal R}$ and total momentum  $-{\bf p}$. Assuming that final state interactions between the struck constituent and the recoiling spectator system be negligible, the energy of the state $|F\rangle$ can be written in the form~\cite{Benhar2000}
\begin{align}
E_F = |{\bf q}| + p_\parallel + E_{\mathcal R} + \ldots \ , 
\end{align}
where $p_\parallel$ is the component of ${\bf p}$ parallel to the momentum transfer and the ellipsis represents terms of order $|{\bf q}|^{-1}$, which are disregarded in the  $|{\bf q}| \to \infty$ limit. As a consequence, conservation of energy in the scattering process requires an energy transfer
\begin{align}
\label{energy:cons} 
\nu & =   |{\bf q}| + p_\parallel + E_{\mathcal R} - M \ ,
\end{align}
with $M$ being the target mass. Equation~\eqref{energy:cons} implies that in the kinematical region corresponding to large momentum transfer the response 
of the target\textemdash which in general depends on {\it both} ${\bf q}$ and $\nu$\textemdash becomes of a function of the single variable  
\begin{align}
{\widetilde y} = \nu - |{\bf q}| \ .
\label{def:ytilde}
\end{align}
Note that in lepton scattering $Q^2 = |{\bf q}|^2 - \nu^2 > 0$, and ${\widetilde y}$ turns out to be negative.

The above derivation shows that the occurrence of scaling results from the onset of a dominant reaction mechanism, independent of 
the underlying dynamics. A slightly different but conceptually equivalent implementation of the above procedure allows one to explain the 
observation of scaling in the variable $y$\textemdash perfectly analogous to ${\widetilde y}$ of Eq.~\eqref{def:ytilde}\textemdash in reactions as different as neutron scattering by liquid helium~\cite{yscaling_helium}  and quasielastic electron-nucleus scattering~\cite{yscaling_3He,RMP}.  In all instances, the scaling variable 
is simply related to the component of the momentum of the struck constituent parallel to the momentum transfer in the target rest frame. 

The authors of Ref.~\cite{Benhar2000} argued that DIS of electrons by protons may also be described using the formalism of many-body theory.
The results of their analysis offer compelling evidence that the proton structure functions extracted from measured cross sections 
corresponding to $|{\bf q}| \gtrsim$ 5 GeV exhibit a striking scaling behaviour when plotted as a function of  ${\widetilde y}$. 
This outcome does not, in fact, come as a surprise, because the same data have 
long been  shown to scale in the Nachtmann variable $\xi$~\cite{Nachtmann}, which reduces to $\xi = -{\widetilde y}/m_N$, $m_N$ being the 
nucleon mass~\cite{Jaffe:1985je}, in the target rest frame. Moreover, because in the $m_N^2/Q^2 \to 0$ limit $\xi$ reduces to the Bjorken scaling variable $x$, 
${\widetilde y}$-scaling turns out to be closely connected to $x$-scaling as well. 
 
The key advantage of using ${\widetilde y}$ as a scaling variable lies in its straightforward interpretation in the rest frame of the target. As pointed out by the authors of
Ref.~\cite{Benhar2000}, the very definition of Eq.~\eqref{def:ytilde} implies that the analysis based on ${\widetilde y}$ allows to unambiguously identify the effect of nuclear binding in DIS. 

To see this, consider that, while in electron-nucleon scattering in free space 
the struck particle is given the entire energy transfer $\nu$, in scattering processes
involving bound nucleons a fraction of the electron energy goes
into the recoiling spectator system; see, e.g., Ref.~\cite{RMP}. Writing the energy transferred to the struck nucleon  in the form
${\widetilde \nu} = \nu - \delta \nu$, it can be easily found that in the $|{\bf p}|/{m_N} \to 0$ limit 
$\delta  \nu ~\approx \langle E_A \rangle$, with the  average nucleon removal energy $\langle E_A \rangle$ being an increasing function of the nuclear mass number $A$. It follows that, compared to that of the deuteron, the structure function per nucleon of a nucleus of mass 
$A$ is shifted towards larger values of ${\widetilde y}$ by an amount~$\sim\langle E_A \rangle$, and the cross section ratio at intermediate ${\widetilde y}$ turns out to be less than unity~\cite{Benhar2000}. 

\begin{figure}[!t]
\includegraphics[scale=0.675]{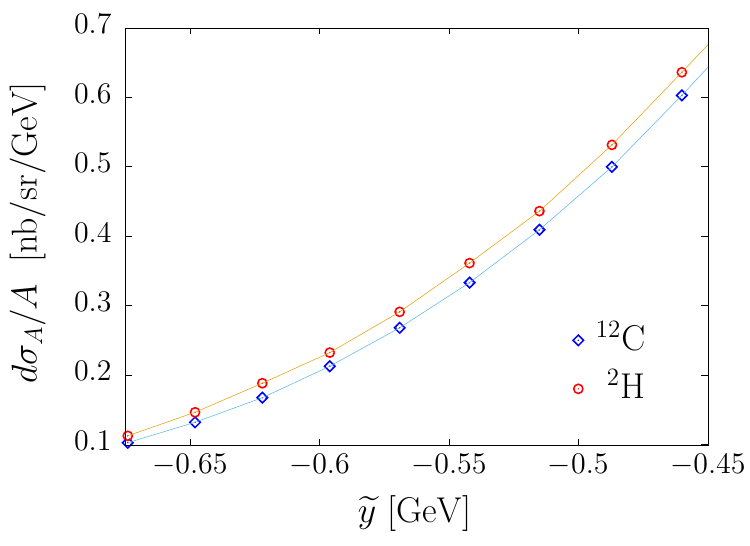}
\vspace*{-0.15in}
\caption{\small Inclusive cross section per nucleon of \isotope[2][]{H} (circles) and \isotope[12][]{C} (diamonds)
at $E_e = 5.766$ GeV and $\theta_e = 40$ deg. 
The JLab data, originally reported by the E03-103 and E02-019 Collaborations in Refs.~\cite{Seely09} and~\cite{Nadia2012}, are taken from the compilation of Ref.~\cite{archive:qe}. Error bars are not visible on the scale of the plot; the lines connecting the data points are meant to guide the eye.\label{xsection_shift}}
\end{figure}

The emergence of the EMC effect from the analysis of the ${\widetilde y}$-dependence of the nuclear structure functions is illustrated in Fig.~\ref{xsection_shift}.
The shift arising from the different nucleon removal energies can be clearly observed by comparing the measured inclusive cross sections per nucleon of \isotope[2][]{H} and \isotope[12][]{C} in the region $-0.7 < {\widetilde y} < -0.45$ GeV, in which $R_A({\widetilde y})<1$; see Fig.~\ref{EMC:ratio} below.


The pattern appearing in Fig.~\ref{xsection_shift} provides the basis of the present study, in which the slopes of the EMC ratios are obtained from their dependence on the scaling variable ${\widetilde y}$. For the sake of illustration, 
the ratio $R_A({\widetilde y})$ corresponding to $A=12$ is displayed in Fig.~\ref{EMC:ratio}.

\begin{figure}[!b]
\includegraphics[scale=0.675]{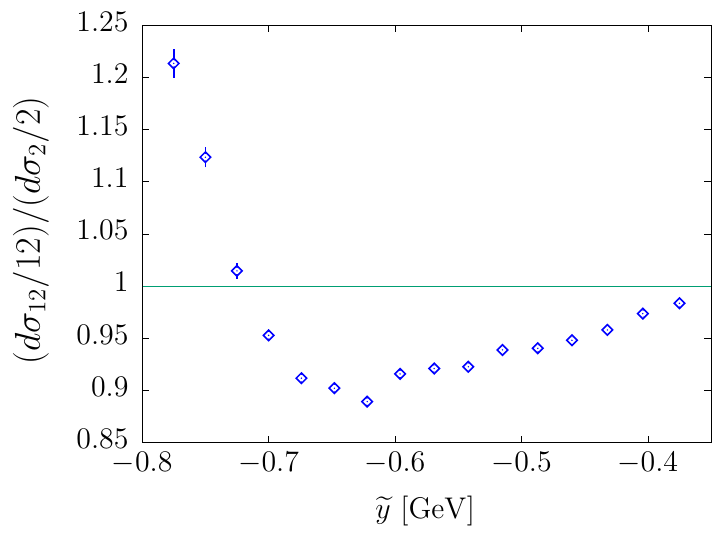}
\vspace*{-0.15in}
\caption{\small Ratio of the inclusive cross section per nucleon of \isotope[2][]{H} and \isotope[12][]{C} 
at $E_e = 5.766$ GeV and  $\theta_e = 40$ deg 
reported in Refs.~\cite{Seely09,Nadia2012} and compiled in Ref.~\cite{archive:qe}. The data are displayed as a function of the scaling variable $\widetilde y$ of Eq.\eqref{def:ytilde}.\label{EMC:ratio}}
\end{figure}

The correlation of the slopes with the nucleon removal energy has been investigated using values of 
$\langle E_A \rangle$ resulting from state-of-the-art quantum many-body calculations, based on realistic models of nuclear dynamics.~\\

\paragraph*{Determination of $\langle E_A \rangle$}
Removing a nucleon of momentum ${\bf p}$ from a nucleus of mass number $A$ leaves the
residual $(A-1)$-nucleon system in a state $| {\mathcal R} \rangle$ which is not, in general, an eigenstate of the Hamiltonian. Its 
energy\textemdash given by $E_{\mathcal R} \approx M_{\mathcal R} = M_{A} - m + E_A$ 
in the ${\bf p}^2/M^2_{\mathcal R} \to 0$ 
limit\textemdash is distributed according to the 
nuclear spectral function, which describes the probability to find a nucleon with removal energy $E_A$~\cite{Pke,Green}. Electron-nucleus scattering experiments have provided convincing evidence that SRC in the target ground state, leading to the excitation of two-particle\textendash two-hole states, 
are associated with the appearance of high-energy contributions, pushing $\langle E_A \rangle$ much beyond the predictions of the nuclear shell model~\cite{RMP}.      

In principle, $\langle E_A \rangle$ should be obtained from the target spectral function, the determination of which involves non trivial difficulties. 
The available models are mostly based on a combination of electron-nucleus scattering data and results of theoretical nuclear matter calculations~\cite{Benhar94,Ankowski}. 
While being adequate to describe a large set of measured inclusive cross sections, however, the spectral functions obtained from this approach 
are limited by both the kinematic range and precision of the input data and the complexity of many-body calculations involving continuum states.
 
More precise estimates of the average nucleon removal energy can be derived from the Galitskii--Migdal--Koltun sum rule~\cite{Koltun,Harada}, allowing 
one to  write $\langle E_A \rangle$ in terms of ground-state expectation values which can be accurately computed using quantum Monte Carlo (QMC) techniques. For realistic Hamiltonian models, which include a potential describing 
irreducible three-nucleon $(3N)$ interactions, the resulting expression is 
\begin{align}
\label{Koltun}
\langle E_{A} \rangle & = \frac{1}{A}\left[\frac{A-2}{A-1}\,\langle T\rangle - \langle V_{3N}\rangle - 2{\mathcal E}_A\right] \ ,
\end{align}
where ${\mathcal E}_A=\langle H \rangle$, $\langle T \rangle$ and $\langle V_{3N}\rangle$ denote the expectation values of the full nuclear Hamiltonian, the 
kinetic energy operator, and the $3N$ potential.  
The factor $(A~-~2)/(A~-~1)$ takes into account the kinetic energy associated with the center-of-mass motion of the recoiling nucleus. 

For all nuclei with  mass number $A \leq 12$ 
we have computed  ${\mathcal E}_A$, $\langle T\rangle$ and $\langle V_{3N}\rangle$ within the Green's Function Monte Carlo (GFMC) scheme~\cite{Carlson:2014vla}. The results corresponding to a Hamiltonian comprising the Argonne $v_{18}$~\cite{Wiringa:1994wb} and Illinois\textendash7~\cite{Pieper:2008rui} two- and three-nucleon potentials, respectively, are listed in Table~\ref{tab:koltun_summary} together with the associated values of 
$\langle E_A \rangle$. We also report 
\isotope[40][]{Ca} results, obtained from the cluster variational Monte Carlo (CVMC) method~\cite{Lonardoni:2017egu} using the Argonne~$v_{18}$ and the Urbana IX~\cite{Pudliner:1995wk} interactions.
It should be noted, however, that CVMC calculations are known to underestimate the ground-state energy of \isotope[40][]{Ca}. 
To gauge the impact of this deficiency on our analysis we have re-evaluated $\langle E_A \rangle$ replacing the CVMC ground-state energy with the 
experimental value, while keeping the kinetic energy and $V_{3N}$ expectation values unchanged. The resulting removal energy of 
\isotope[40][]{Ca} is given in the  bottom line of Table~\ref{tab:koltun_summary}. The validity of our procedure is supported by a comparison between
the average removal energy of \isotope[16][]{O} obtained from CVMC and the one computed using the highly accurate  auxiliary-field diffusion Monte Carlo (AFDMC) method~\cite{Schmidt:1999lik}; see End Matter for details.~\\


\begin{table}[t]
\centering
\caption{\small
Ground-state expectation values per nucleon and associated removal energies, obtained from Eq.~\eqref{Koltun}.
 The results corresponding to $A \leq 12$ have been obtained from GFMC calculations using the Argonne~$v_{18}$ and Illinois\textendash7 potentials; 
 for \isotope[40][]{Ca} the energies have been computed using the CVMC approach and replacing the Illinois\textendash7 potential with the Urbana IX model.
The bottom line reports value of $\langle E_A \rangle$ corrected to take into account the uncertainty associated with the variational method; see text for details. All energies are in MeV. 
}
\label{tab:koltun_summary}
\vspace*{.10in}
\begin{tabular}{lccccc}
\hline\hline
Nucleus & \ \ ${\mathcal E}_A/A$ & \ \ $\langle T\rangle/A$ & \ \ \ $\langle V_{3N}\rangle/A$ & \ \ $\langle E_A \rangle$ \\
\hline
\isotope[2][]{H}    & $-1.113$ & \ \ \ $9.905$ &  \ \  $0.000$       & \ $2.225$ \\
\isotope[3][]{He}  & $-2.573$ & \ \ $16.967$ & $-0.367$  & \ $13.997$ \\
\isotope[4][]{He}  & $-7.105$ & \ \ $28.000$ & $-1.550$  & \ $34.427$ \\
\isotope[9][]{Be}  & $-6.433$ &  \ \ $31.333$ & $-1.978$  & \ $42.261$ \\
\isotope[10][]{B}  & $-6.470$ &  \ \ $33.900$ & $-2.550$  & \ $45.623$ \\
\isotope[11][]{B}  & $-6.691$ &  \ \ $33.545$ & $-2.282$ & \ $45.855$ \\
\isotope[12][]{C}  & $-7.775$ & \ \ $36.417$ & $-2.933$  & \ $51.589$ \\
\hline
\multirow{2}{*}{\isotope[40][]{Ca}}  & $-4.920$ & \ \ $30.970$ & $-0.230$ & \ $40.246$ \\
 & $-8.520$ &  \ \ $30.970$ & $-0.230$  & \ $47.446$ \\
\hline\hline
\end{tabular}
\end{table}

\paragraph*{Data analysis and results} 

The slopes of the EMC ratios corresponding to the JLab data reported in Refs.~\cite{Seely09,Arrington12,Arrington21} 
have been obtained from the inclusive cross sections of \isotope[2][]{H}, \isotope[3][]{He}, \isotope[4][]{He}, \isotope[9][]{Be} 
and \isotope[12][]{C} compiled in Ref.~\cite{archive:qe}. Our analysis covers the same kinematic region studied by the authors 
of Ref.~\cite{Seely09}, corresponding to $0.35 < x < 0.7$. 
For non-isospin symmetric nuclei, the effect of neutron excess has 
been taken into account using the procedure described in Ref.~\cite{Arrington21}. 
The EMC ratios of \isotope[8][]{Be}, \isotope[10][]{B}, \isotope[11][]{B} and \isotope[12][]{C} recently reported by 
Karki {\it et al.}~\cite{Karki2023}\textemdash obtained from measurements performed in JLab Hall C using a 10.6 GeV electron beam\textemdash  
have been also included in our study. On the other hand, the data collected in Hall B by the CLAS collaboration~\cite{CLAS:2019} 
were not available to us in a form suitable to consider their dependence on ${\widetilde y}$. 

Finally, we have analysed the $Q^2$-averaged EMC ratios of \isotope[4][]{He}, 
\isotope[9][]{Be}, \isotope[12][]{C} and  \isotope[40][]{Ca} reported by the SLAC E139 experiment~\cite{Gomez94}, as well as the one  
corresponding to isospin-symmetric nuclear matter (SNM) at equilibrium density, obtained by the authors of Ref.~\cite{NM_EMC} performing a $A \to \infty$ extrapolation based on 
the inclusive cross sections of \isotope[4][]{He}, \isotope[12][]{C}, \isotope[27][]{Al}, \isotope[56][]{Fe} and \isotope[197][]{Au} measured at SLAC~\cite{NM_response}. 

\begin{figure}[htb!]
\includegraphics[scale=0.675]{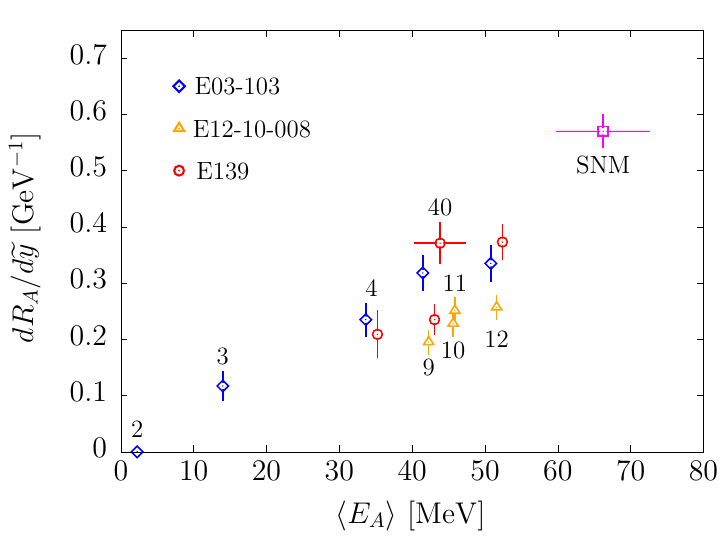}
\vspace*{-0.15in}
\caption{\small Slopes of the EMC ratios in the ${\widetilde y}$ region corresponding to $0.35 < x < 0.7$. 
The results, labeled according to the target mass number $A$, have been obtained from data 
collected by the E03-103 and E12-10-008 experiments at JLab and the E139 experiment at SLAC.
For clarity, the E139 and E03-103 points corresponding to $A=$ 4, 9 and 12 are offset by $\pm 0.8$ MeV, respectively. The nuclear matter result, represented by the square, has been obtained from the 
EMC ratio reported in Ref.~\cite{NM_EMC}. \label{EMC:shift}}
\end{figure}

The results of our work are summarised in Fig.~\ref{EMC:shift}, displaying the the slopes of the EMC ratios 
obtained from the data collected by JLab experiments E03-103~\cite{Arrington21} and E12-10-008~\cite{Karki2023} and SLAC 
experiment E139~\cite{Gomez94}. For clarity, the E139 and E03-103 points corresponding to $A =$ 4, 9, and 12 are offset by $\pm 0.8$ MeV, respectively. The nuclear matter result, represented by the square, has been obtained from the EMC ratio reported in Ref.~\cite{NM_EMC}. The horizontal error bars associated with  
the removal energies of \isotope[40][]{Ca} and SNM account for the uncertainty arising from the variational determination of $\langle E_A \rangle$ 
discussed  above. In the case of SNM, the ground-state energy per nucleon obtained by the authors of Ref.~\cite{AP} using the variational approach and 
the formalism of correlated basis functions (CBF) turns out to be ${\mathcal E}_0 = -10.07 \pm 0.82$ MeV, to be compared with the 
empirical value ${\mathcal E}_0 \simeq -16$ MeV. 
The removal energies obtained from Eq.~\eqref{Koltun} combining the calculated and empirical values 
of  ${\mathcal E}_0$ with the expectation values $\langle T \rangle/A$ and $\langle V_{3N}\rangle/A$ reported in Ref.~\cite{AP} are $\langle E_{\rm SNM} \rangle$~=~60.3 and 72.2 
MeV, respectively. It is remarkable that the average removal energy of SNM calculated from the spectral function of Ref.~\cite{Pke}, 
computed in CBF perturbation theory, is within less than 3\% of the one obtained from Eq.~\eqref{Koltun} using the results of 
Ref.~\cite{AP}.

Overall, the pattern observed in Fig.~\ref{EMC:shift} clearly hints at a linear relationship between the slopes of the EMC ratios and the average nucleon removal energies, although the results obtained from the E12-10-008 data turn out to lie somewhat below those corresponding to the other experiments. This discrepancy, which also emerged from the standard analysis of Karki et al.~\cite{Karki2023}, is, in fact, more visible in our results. To understand this feature, consider that, in general, the 
deviation between the slopes obtained by analysing the dependence of $R_A$ on $x$ and ${\widetilde y}$
is a decreasing function of $Q^2$. In the present study, the effect of the $x \to {\widetilde y}$ transformation is largest for the E03-103 data, and the corresponding slopes turn out to be enhanced by a noticeable amount.    

It should be kept in mind, however, that, even in the regime in which 
of ${\widetilde y} \approx - m_N x$ and the slopes of the EMC ratios are little affected by the transformation, the scaling variables ${\widetilde y}$ and $x$ have a profoundly different physical interpretation.  
 
\paragraph*{Conclusions}

The general picture emerging from our study 
appears to be consistent with the prediction of a linear dependence of the EMC effect on the average nucleon removal energy, thereby supporting the validity of the description of nuclear DIS data in terms of the scaling variable ${\widetilde y}$ derived in Ref.~\cite{Benhar2000}. 
The identification of $\langle E_A \rangle$ as the primary factor determining the EMC ratio at intermediate ${\widetilde y}$
is also strengthened by the measurements reported in Ref.~\cite{Karki2023}, showing that the 
\isotope[10][]{B} and \isotope[11][]{B} ratios are consistent within experimental error. This 
observation simply reflects the fact that the corresponding removal energies, listed in Table~\ref{tab:koltun_summary}, only differ by half of a percent. 


The approach employed in our analysis allows to clarify the somewhat elusive role of SRC.
Semi-inclusive electron-nucleus scattering experiments 
have long established that high-momentum nucleons belonging to strongly correlated pairs 
have large removal energies, originating from excitation of the recoiling spectator system to continuum states~\cite{Marchand,Vanleeuwe,Rohe}.   
These data demonstrate that SRC significantly affect the average nucleon 
removal energy, which in turn determines the size of the EMC effect.
 
A quantitative assessment of the dependence of $\langle E_A \rangle$ on SRC can be obtained by
considering the case of SNM at 
equilibrium density. The result of a direct calculation, performed using the spectral function of Ref.~\cite{Pke}, shows that the strength located at momenta larger than the Fermi 
momentum $p_F \approx 260$~MeV\textemdash which 
would vanish identically in the absence of ground-state correlations\textemdash
contributes 37\% of the average removal energy. 

It should be noted that the linear dependence of the EMC effect on $\langle E_A \rangle$ illustrated in Fig.~\ref{EMC:shift} is, in fact, largely 
independent of the microscopic description of nuclear dynamics. On the other hand, nuclear dynamics determines the size of the effect,  
measured by the slope of the EMC ratio. The results obtained by repeating our analysis with an alternative Hamiltonian model, based on chiral effective field theory ($\chi$EFT), are discussed in the End Matter. 


The results reported in this Letter show that the analysis based on the variable $\widetilde y$\textemdash which, unlike the Bjorken variable $x$, has a straightforward physical interpretation in the rest frame of the target nucleus\textemdash 
provides a largely model-independent explanation of both the origin and the $A$-dependence of the EMC effect in the kinematic region
in which nuclear binding plays a leading role.~\\
 
\paragraph*{Acknowledgments} This Letter is dedicated to the memory of our friend and colleague Ingo Sick,
whose thoughts and advice lastingly inspired our work. We would like to thank Diego Lonardoni and Maria Piarulli for providing the AFDMC and GFMC energy values computed using $\chi$EFT potentials. Useful comments by J. Arrington are also gratefully acknowledged. The work of A. L. is supported by the U.S. Department of Energy, Office of Science, Office of Nuclear Physics, under contract DE-AC02-06CH11357, by the DOE Early Career Research Program, by the Office of Advanced Scientific Computing Research, Scientific Discovery through Advanced Computing (SciDAC) NUCLEI program, and by grant PID2023-147458NB-C21 funded by MCIN/AEI/10.13039/501100011033, and by the European Union. The work of O. B. is supported by the Italian National Institute for Nuclear Research (INFN) under grant DOT4.


\begin{thebibliography}{55}%
\makeatletter
\providecommand \@ifxundefined [1]{%
 \@ifx{#1\undefined}
}%
\providecommand \@ifnum [1]{%
 \ifnum #1\expandafter \@firstoftwo
 \else \expandafter \@secondoftwo
 \fi
}%
\providecommand \@ifx [1]{%
 \ifx #1\expandafter \@firstoftwo
 \else \expandafter \@secondoftwo
 \fi
}%
\providecommand \natexlab [1]{#1}%
\providecommand \enquote  [1]{``#1''}%
\providecommand \bibnamefont  [1]{#1}%
\providecommand \bibfnamefont [1]{#1}%
\providecommand \citenamefont [1]{#1}%
\providecommand \href@noop [0]{\@secondoftwo}%
\providecommand \href [0]{\begingroup \@sanitize@url \@href}%
\providecommand \@href[1]{\@@startlink{#1}\@@href}%
\providecommand \@@href[1]{\endgroup#1\@@endlink}%
\providecommand \@sanitize@url [0]{\catcode `\\12\catcode `\$12\catcode
  `\&12\catcode `\#12\catcode `\^12\catcode `\_12\catcode `\%12\relax}%
\providecommand \@@startlink[1]{}%
\providecommand \@@endlink[0]{}%
\providecommand \url  [0]{\begingroup\@sanitize@url \@url }%
\providecommand \@url [1]{\endgroup\@href {#1}{\urlprefix }}%
\providecommand \urlprefix  [0]{URL }%
\providecommand \Eprint [0]{\href }%
\providecommand \doibase [0]{https://doi.org/}%
\providecommand \selectlanguage [0]{\@gobble}%
\providecommand \bibinfo  [0]{\@secondoftwo}%
\providecommand \bibfield  [0]{\@secondoftwo}%
\providecommand \translation [1]{[#1]}%
\providecommand \BibitemOpen [0]{}%
\providecommand \bibitemStop [0]{}%
\providecommand \bibitemNoStop [0]{.\EOS\space}%
\providecommand \EOS [0]{\spacefactor3000\relax}%
\providecommand \BibitemShut  [1]{\csname bibitem#1\endcsname}%
\let\auto@bib@innerbib\@empty
\bibitem [{CER(1982)}]{CERN_courier_1982}%
  \BibitemOpen
  \href@noop {} {\bibfield  {journal} {\bibinfo  {journal} {CERN Courier}\
  }\textbf {\bibinfo {volume} {9}},\ \bibinfo {pages} {362} (\bibinfo {year}
  {1982})}\BibitemShut {NoStop}%
\bibitem [{\citenamefont {{J. J. Aubert, {\it et al.} (EMC
  Collaboration)}}(1983)}]{EMC_1983}%
  \BibitemOpen
  \bibfield  {author} {\bibinfo {author} {\bibnamefont {{J. J. Aubert, {\it et
  al.} (EMC Collaboration)}}},\ }\href
  {https://doi.org/10.1016/0370-2693(83)90437-9} {\bibfield  {journal}
  {\bibinfo  {journal} {Phys. Lett. B}\ }\textbf {\bibinfo {volume} {123}},\
  \bibinfo {pages} {275} (\bibinfo {year} {1983})}\BibitemShut {NoStop}%
\bibitem [{\citenamefont {{G. Bari, {\it et al.} (BCDMS
  Collaboration)}}(1985)}]{Bari85}%
  \BibitemOpen
  \bibfield  {author} {\bibinfo {author} {\bibnamefont {{G. Bari, {\it et al.}
  (BCDMS Collaboration)}}},\ }\href
  {https://doi.org/https://doi.org/10.1016/0370-2693(85)90238-2} {\bibfield
  {journal} {\bibinfo  {journal} {Phys. Lett. B}\ }\textbf {\bibinfo {volume}
  {163}},\ \bibinfo {pages} {282} (\bibinfo {year} {1985})}\BibitemShut
  {NoStop}%
\bibitem [{\citenamefont {{J. Ashman, {\it et al.} (EMC
  Collaboration)}}(1988)}]{Ashman88}%
  \BibitemOpen
  \bibfield  {author} {\bibinfo {author} {\bibnamefont {{J. Ashman, {\it et
  al.} (EMC Collaboration)}}},\ }\href
  {https://doi.org/https://doi.org/10.1016/0370-2693(88)91872-2} {\bibfield
  {journal} {\bibinfo  {journal} {Phys. Lett. B}\ }\textbf {\bibinfo {volume}
  {202}},\ \bibinfo {pages} {603} (\bibinfo {year} {1988})}\BibitemShut
  {NoStop}%
\bibitem [{\citenamefont {Gomez}\ \emph {et~al.}(1994)\citenamefont {Gomez},
  \citenamefont {Arnold}, \citenamefont {Bosted}, \citenamefont {Chang},
  \citenamefont {Katramatou}, \citenamefont {Petratos}, \citenamefont {Rahbar},
  \citenamefont {Rock}, \citenamefont {Sill}, \citenamefont {Szalata},
  \citenamefont {Bodek}, \citenamefont {Giokaris}, \citenamefont {Sherden},
  \citenamefont {Mecking},\ and\ \citenamefont {Lombard-Nelsen}}]{Gomez94}%
  \BibitemOpen
  \bibfield  {author} {\bibinfo {author} {\bibfnamefont {J.}~\bibnamefont
  {Gomez}}, \bibinfo {author} {\bibfnamefont {R.~G.}\ \bibnamefont {Arnold}},
  \bibinfo {author} {\bibfnamefont {P.~E.}\ \bibnamefont {Bosted}}, \bibinfo
  {author} {\bibfnamefont {C.~C.}\ \bibnamefont {Chang}}, \bibinfo {author}
  {\bibfnamefont {A.~T.}\ \bibnamefont {Katramatou}}, \bibinfo {author}
  {\bibfnamefont {G.~G.}\ \bibnamefont {Petratos}}, \bibinfo {author}
  {\bibfnamefont {A.~A.}\ \bibnamefont {Rahbar}}, \bibinfo {author}
  {\bibfnamefont {S.~E.}\ \bibnamefont {Rock}}, \bibinfo {author}
  {\bibfnamefont {A.~F.}\ \bibnamefont {Sill}}, \bibinfo {author}
  {\bibfnamefont {Z.~M.}\ \bibnamefont {Szalata}}, \bibinfo {author}
  {\bibfnamefont {A.}~\bibnamefont {Bodek}}, \bibinfo {author} {\bibfnamefont
  {N.}~\bibnamefont {Giokaris}}, \bibinfo {author} {\bibfnamefont {D.~J.}\
  \bibnamefont {Sherden}}, \bibinfo {author} {\bibfnamefont {B.~A.}\
  \bibnamefont {Mecking}},\ and\ \bibinfo {author} {\bibfnamefont {R.~M.}\
  \bibnamefont {Lombard-Nelsen}},\ }\href
  {https://doi.org/10.1103/PhysRevD.49.4348} {\bibfield  {journal} {\bibinfo
  {journal} {Phys. Rev. D}\ }\textbf {\bibinfo {volume} {49}},\ \bibinfo
  {pages} {4348} (\bibinfo {year} {1994})}\BibitemShut {NoStop}%
\bibitem [{\citenamefont {{K. Ackerstaff, {\it et al.} (HERMES
  Collaboration)}}(2000)}]{Ackerstaff2000}%
  \BibitemOpen
  \bibfield  {author} {\bibinfo {author} {\bibnamefont {{K. Ackerstaff, {\it et
  al.} (HERMES Collaboration)}}},\ }\href
  {https://doi.org/https://doi.org/10.1016/S0370-2693(99)01493-8} {\bibfield
  {journal} {\bibinfo  {journal} {Phys. Lett. B}\ }\textbf {\bibinfo {volume}
  {475}},\ \bibinfo {pages} {386} (\bibinfo {year} {2000})}\BibitemShut
  {NoStop}%
\bibitem [{\citenamefont {{A. Airapetian, {\it et al.}, (HERMES
  Collaboration)}}(2003)}]{Airapetian2003}%
  \BibitemOpen
  \bibfield  {author} {\bibinfo {author} {\bibnamefont {{A. Airapetian, {\it et
  al.}, (HERMES Collaboration)}}},\ }\href
  {https://doi.org/https://doi.org/10.1016/j.physletb.2003.06.044} {\bibfield
  {journal} {\bibinfo  {journal} {Phys. Lett. B}\ }\textbf {\bibinfo {volume}
  {567}},\ \bibinfo {pages} {339} (\bibinfo {year} {2003})}\BibitemShut
  {NoStop}%
\bibitem [{\citenamefont {{J. Seely, {\it et al.}}}(2009)}]{Seely09}%
  \BibitemOpen
  \bibfield  {author} {\bibinfo {author} {\bibnamefont {{J. Seely, {\it et
  al.}}}},\ }\href {https://doi.org/10.1103/PhysRevLett.103.202301} {\bibfield
  {journal} {\bibinfo  {journal} {Phys. Rev. Lett.}\ }\textbf {\bibinfo
  {volume} {103}},\ \bibinfo {pages} {202301} (\bibinfo {year}
  {2009})}\BibitemShut {NoStop}%
\bibitem [{\citenamefont {Arrington}\ \emph {et~al.}(2012)\citenamefont
  {Arrington}, \citenamefont {Daniel}, \citenamefont {Day}, \citenamefont
  {Fomin}, \citenamefont {Gaskell},\ and\ \citenamefont
  {Solvignon}}]{Arrington12}%
  \BibitemOpen
  \bibfield  {author} {\bibinfo {author} {\bibfnamefont {J.}~\bibnamefont
  {Arrington}}, \bibinfo {author} {\bibfnamefont {A.}~\bibnamefont {Daniel}},
  \bibinfo {author} {\bibfnamefont {D.~B.}\ \bibnamefont {Day}}, \bibinfo
  {author} {\bibfnamefont {N.}~\bibnamefont {Fomin}}, \bibinfo {author}
  {\bibfnamefont {D.}~\bibnamefont {Gaskell}},\ and\ \bibinfo {author}
  {\bibfnamefont {P.}~\bibnamefont {Solvignon}},\ }\href
  {https://doi.org/10.1103/PhysRevC.86.065204} {\bibfield  {journal} {\bibinfo
  {journal} {Phys. Rev. C}\ }\textbf {\bibinfo {volume} {86}},\ \bibinfo
  {pages} {065204} (\bibinfo {year} {2012})}\BibitemShut {NoStop}%
\bibitem [{\citenamefont {Schmookler}\ \emph {et~al.}(2019)\citenamefont
  {Schmookler} \emph {et~al.}}]{CLAS:2019}%
  \BibitemOpen
  \bibfield  {author} {\bibinfo {author} {\bibfnamefont {B.}~\bibnamefont
  {Schmookler}} \emph {et~al.} (\bibinfo {collaboration} {CLAS
  Collaboration}),\ }\href {https://doi.org/10.1038/s41586-019-0925-9}
  {\bibfield  {journal} {\bibinfo  {journal} {Nature}\ }\textbf {\bibinfo
  {volume} {566}},\ \bibinfo {pages} {354} (\bibinfo {year}
  {2019})}\BibitemShut {NoStop}%
\bibitem [{\citenamefont {{J. Arrington, {\it et al.}}}(2021)}]{Arrington21}%
  \BibitemOpen
  \bibfield  {author} {\bibinfo {author} {\bibnamefont {{J. Arrington, {\it et
  al.}}}},\ }\href {https://doi.org/10.1103/PhysRevC.104.065203} {\bibfield
  {journal} {\bibinfo  {journal} {Phys. Rev. C}\ }\textbf {\bibinfo {volume}
  {104}},\ \bibinfo {pages} {065203} (\bibinfo {year} {2021})}\BibitemShut
  {NoStop}%
\bibitem [{\citenamefont {{A. Karki, {\it et al.}}}(2023)}]{Karki2023}%
  \BibitemOpen
  \bibfield  {author} {\bibinfo {author} {\bibnamefont {{A. Karki, {\it et
  al.}}}} (\bibinfo {collaboration} {Hall C Collaboration}),\ }\href
  {https://doi.org/10.1103/PhysRevC.108.035201} {\bibfield  {journal} {\bibinfo
   {journal} {Phys. Rev. C}\ }\textbf {\bibinfo {volume} {108}},\ \bibinfo
  {pages} {035201} (\bibinfo {year} {2023})}\BibitemShut {NoStop}%
\bibitem [{\citenamefont {{D. W. Higinbotham, G. A. Miller, O. Hen, and K.
  Rith}}(2013)}]{CERN_courier_2013}%
  \BibitemOpen
  \bibfield  {author} {\bibinfo {author} {\bibnamefont {{D. W. Higinbotham, G.
  A. Miller, O. Hen, and K. Rith}}},\ }\href@noop {} {\bibfield  {journal}
  {\bibinfo  {journal} {CERN Courier}\ }\textbf {\bibinfo {volume} {53}},\
  \bibinfo {pages} {24} (\bibinfo {year} {2013})}\BibitemShut {NoStop}%
\bibitem [{\citenamefont {Arneodo}(1994)}]{Arneodo94}%
  \BibitemOpen
  \bibfield  {author} {\bibinfo {author} {\bibfnamefont {M.}~\bibnamefont
  {Arneodo}},\ }\href
  {https://doi.org/https://doi.org/10.1016/0370-1573(94)90048-5} {\bibfield
  {journal} {\bibinfo  {journal} {Phys. Rep.}\ }\textbf {\bibinfo {volume}
  {240}},\ \bibinfo {pages} {301} (\bibinfo {year} {1994})}\BibitemShut
  {NoStop}%
\bibitem [{\citenamefont {Geesaman}\ \emph {et~al.}(1995)\citenamefont
  {Geesaman}, \citenamefont {Saito},\ and\ \citenamefont
  {Thomas}}]{Geesaman95}%
  \BibitemOpen
  \bibfield  {author} {\bibinfo {author} {\bibfnamefont {D.~F.}\ \bibnamefont
  {Geesaman}}, \bibinfo {author} {\bibfnamefont {K.}~\bibnamefont {Saito}},\
  and\ \bibinfo {author} {\bibfnamefont {A.~W.}\ \bibnamefont {Thomas}},\
  }\href {https://doi.org/10.1146/annurev.ns.45.120195.002005} {\bibfield
  {journal} {\bibinfo  {journal} {Ann. Rev. Nucl. Part. Sci.}\ }\textbf
  {\bibinfo {volume} {45}},\ \bibinfo {pages} {337} (\bibinfo {year}
  {1995})}\BibitemShut {NoStop}%
\bibitem [{\citenamefont {{P. R. Norton}}(2003)}]{Norton03}%
  \BibitemOpen
  \bibfield  {author} {\bibinfo {author} {\bibnamefont {{P. R. Norton}}},\
  }\href {https://doi.org/10.1088/0034-4885/66/8/201} {\bibfield  {journal}
  {\bibinfo  {journal} {Rept. Prog. Phys.}\ }\textbf {\bibinfo {volume} {66}},\
  \bibinfo {pages} {1253} (\bibinfo {year} {2003})}\BibitemShut {NoStop}%
\bibitem [{\citenamefont {Malace}\ \emph {et~al.}(2014)\citenamefont {Malace},
  \citenamefont {Gaskell}, \citenamefont {Higinbotham},\ and\ \citenamefont
  {Cloet}}]{Malace:2014}%
  \BibitemOpen
  \bibfield  {author} {\bibinfo {author} {\bibfnamefont {S.}~\bibnamefont
  {Malace}}, \bibinfo {author} {\bibfnamefont {D.}~\bibnamefont {Gaskell}},
  \bibinfo {author} {\bibfnamefont {D.~W.}\ \bibnamefont {Higinbotham}},\ and\
  \bibinfo {author} {\bibfnamefont {I.}~\bibnamefont {Cloet}},\ }\href
  {https://doi.org/10.1142/S0218301314300136} {\bibfield  {journal} {\bibinfo
  {journal} {Int. J. Mod. Phys. E}\ }\textbf {\bibinfo {volume} {23}},\
  \bibinfo {pages} {1430013} (\bibinfo {year} {2014})}\BibitemShut {NoStop}%
\bibitem [{\citenamefont {Ciofi~degli Atti}\ and\ \citenamefont
  {Liuti}(1991)}]{Ciofi91}%
  \BibitemOpen
  \bibfield  {author} {\bibinfo {author} {\bibfnamefont {C.}~\bibnamefont
  {Ciofi~degli Atti}}\ and\ \bibinfo {author} {\bibfnamefont {S.}~\bibnamefont
  {Liuti}},\ }\href {https://doi.org/10.1103/PhysRevC.44.R1269} {\bibfield
  {journal} {\bibinfo  {journal} {Phys. Rev. C}\ }\textbf {\bibinfo {volume}
  {44}},\ \bibinfo {pages} {R1269} (\bibinfo {year} {1991})}\BibitemShut
  {NoStop}%
\bibitem [{\citenamefont {Benhar}\ \emph {et~al.}(1997)\citenamefont {Benhar},
  \citenamefont {Pandharipande},\ and\ \citenamefont {Sick}}]{Benhar97}%
  \BibitemOpen
  \bibfield  {author} {\bibinfo {author} {\bibfnamefont {O.}~\bibnamefont
  {Benhar}}, \bibinfo {author} {\bibfnamefont {V.~R.}\ \bibnamefont
  {Pandharipande}},\ and\ \bibinfo {author} {\bibfnamefont {I.}~\bibnamefont
  {Sick}},\ }\href {https://doi.org/10.1016/S0370-2693(97)00943-X} {\bibfield
  {journal} {\bibinfo  {journal} {Phys. Lett. B}\ }\textbf {\bibinfo {volume}
  {410}},\ \bibinfo {pages} {79} (\bibinfo {year} {1997})}\BibitemShut
  {NoStop}%
\bibitem [{\citenamefont {Weinstein}\ \emph {et~al.}(2011)\citenamefont
  {Weinstein}, \citenamefont {Piasetzky}, \citenamefont {Higinbotham},
  \citenamefont {Gomez}, \citenamefont {Hen},\ and\ \citenamefont
  {Shneor}}]{Weinstein11}%
  \BibitemOpen
  \bibfield  {author} {\bibinfo {author} {\bibfnamefont {L.~B.}\ \bibnamefont
  {Weinstein}}, \bibinfo {author} {\bibfnamefont {E.}~\bibnamefont
  {Piasetzky}}, \bibinfo {author} {\bibfnamefont {D.~W.}\ \bibnamefont
  {Higinbotham}}, \bibinfo {author} {\bibfnamefont {J.}~\bibnamefont {Gomez}},
  \bibinfo {author} {\bibfnamefont {O.}~\bibnamefont {Hen}},\ and\ \bibinfo
  {author} {\bibfnamefont {R.}~\bibnamefont {Shneor}},\ }\href
  {https://doi.org/10.1103/PhysRevLett.106.052301} {\bibfield  {journal}
  {\bibinfo  {journal} {Phys. Rev. Lett.}\ }\textbf {\bibinfo {volume} {106}},\
  \bibinfo {pages} {052301} (\bibinfo {year} {2011})}\BibitemShut {NoStop}%
\bibitem [{\citenamefont {{O. Hen, E. Piasetzky, and L. B.
  Weinstein}}(2012)}]{Hen12}%
  \BibitemOpen
  \bibfield  {author} {\bibinfo {author} {\bibnamefont {{O. Hen, E. Piasetzky,
  and L. B. Weinstein}}},\ }\href {https://doi.org/10.1103/PhysRevC.85.047301}
  {\bibfield  {journal} {\bibinfo  {journal} {Phys. Rev. C}\ }\textbf {\bibinfo
  {volume} {85}},\ \bibinfo {pages} {047301} (\bibinfo {year}
  {2012})}\BibitemShut {NoStop}%
\bibitem [{\citenamefont {{N. Fomin, {\it et al.}}}(2012)}]{Nadia2012}%
  \BibitemOpen
  \bibfield  {author} {\bibinfo {author} {\bibnamefont {{N. Fomin, {\it et
  al.}}}},\ }\href {https://doi.org/10.1103/PhysRevLett.108.092502} {\bibfield
  {journal} {\bibinfo  {journal} {Phys. Rev. Lett.}\ }\textbf {\bibinfo
  {volume} {108}},\ \bibinfo {pages} {092502} (\bibinfo {year}
  {2012})}\BibitemShut {NoStop}%
\bibitem [{\citenamefont {Arrington}\ and\ \citenamefont
  {Fomin}(2019)}]{Arrington2019}%
  \BibitemOpen
  \bibfield  {author} {\bibinfo {author} {\bibfnamefont {J.}~\bibnamefont
  {Arrington}}\ and\ \bibinfo {author} {\bibfnamefont {N.}~\bibnamefont
  {Fomin}},\ }\href {https://doi.org/10.1103/PhysRevLett.123.042501} {\bibfield
   {journal} {\bibinfo  {journal} {Phys. Rev. Lett.}\ }\textbf {\bibinfo
  {volume} {123}},\ \bibinfo {pages} {042501} (\bibinfo {year}
  {2019})}\BibitemShut {NoStop}%
\bibitem [{\citenamefont {Fomin}\ \emph {et~al.}(2026)\citenamefont {Fomin}
  \emph {et~al.}}]{Fomin:2026swt}%
  \BibitemOpen
  \bibfield  {author} {\bibinfo {author} {\bibfnamefont {N.}~\bibnamefont
  {Fomin}} \emph {et~al.},\ }\href@noop {} {\bibinfo {title} {{Long Range
  Outlook for Short-Range Correlations}}} (\bibinfo {year} {2026}),\ \Eprint
  {https://arxiv.org/abs/2601.09568} {arXiv:2601.09568 [nucl-ex]} \BibitemShut
  {NoStop}%
\bibitem [{\citenamefont {Benhar}\ and\ \citenamefont
  {Sick}(2012)}]{Benhar2012}%
  \BibitemOpen
  \bibfield  {author} {\bibinfo {author} {\bibfnamefont {O.}~\bibnamefont
  {Benhar}}\ and\ \bibinfo {author} {\bibfnamefont {I.}~\bibnamefont {Sick}},\
  }\href@noop {} {\  (\bibinfo {year} {2012})},\ \Eprint
  {https://arxiv.org/abs/1207.4595} {arXiv:1207.4595 [nucl-th]} \BibitemShut
  {NoStop}%
\bibitem [{\citenamefont {Benhar}\ \emph {et~al.}(2000)\citenamefont {Benhar},
  \citenamefont {Pandharipande},\ and\ \citenamefont {Sick}}]{Benhar2000}%
  \BibitemOpen
  \bibfield  {author} {\bibinfo {author} {\bibfnamefont {O.}~\bibnamefont
  {Benhar}}, \bibinfo {author} {\bibfnamefont {V.~R.}\ \bibnamefont
  {Pandharipande}},\ and\ \bibinfo {author} {\bibfnamefont {I.}~\bibnamefont
  {Sick}},\ }\href
  {https://doi.org/https://doi.org/10.1016/S0370-2693(00)00909-6} {\bibfield
  {journal} {\bibinfo  {journal} {Phys. Lett. B}\ }\textbf {\bibinfo {volume}
  {489}},\ \bibinfo {pages} {131} (\bibinfo {year} {2000})}\BibitemShut
  {NoStop}%
\bibitem [{\citenamefont {Azuah}\ \emph {et~al.}(1997)\citenamefont {Azuah},
  \citenamefont {Stirling}, \citenamefont {Glyde}, \citenamefont {Boninsegni},
  \citenamefont {Sokol},\ and\ \citenamefont {Bennington}}]{yscaling_helium}%
  \BibitemOpen
  \bibfield  {author} {\bibinfo {author} {\bibfnamefont {R.~T.}\ \bibnamefont
  {Azuah}}, \bibinfo {author} {\bibfnamefont {W.~G.}\ \bibnamefont {Stirling}},
  \bibinfo {author} {\bibfnamefont {H.~R.}\ \bibnamefont {Glyde}}, \bibinfo
  {author} {\bibfnamefont {M.}~\bibnamefont {Boninsegni}}, \bibinfo {author}
  {\bibfnamefont {P.~E.}\ \bibnamefont {Sokol}},\ and\ \bibinfo {author}
  {\bibfnamefont {S.~M.}\ \bibnamefont {Bennington}},\ }\href
  {https://doi.org/10.1103/PhysRevB.56.14620} {\bibfield  {journal} {\bibinfo
  {journal} {Phys. Rev. B}\ }\textbf {\bibinfo {volume} {56}},\ \bibinfo
  {pages} {14620} (\bibinfo {year} {1997})}\BibitemShut {NoStop}%
\bibitem [{\citenamefont {Sick}\ \emph {et~al.}(1980)\citenamefont {Sick},
  \citenamefont {Day},\ and\ \citenamefont {McCarthy}}]{yscaling_3He}%
  \BibitemOpen
  \bibfield  {author} {\bibinfo {author} {\bibfnamefont {I.}~\bibnamefont
  {Sick}}, \bibinfo {author} {\bibfnamefont {D.}~\bibnamefont {Day}},\ and\
  \bibinfo {author} {\bibfnamefont {J.~S.}\ \bibnamefont {McCarthy}},\ }\href
  {https://doi.org/10.1103/PhysRevLett.45.871} {\bibfield  {journal} {\bibinfo
  {journal} {Phys. Rev. Lett.}\ }\textbf {\bibinfo {volume} {45}},\ \bibinfo
  {pages} {871} (\bibinfo {year} {1980})}\BibitemShut {NoStop}%
\bibitem [{\citenamefont {Benhar}\ \emph {et~al.}(2008)\citenamefont {Benhar},
  \citenamefont {Day},\ and\ \citenamefont {Sick}}]{RMP}%
  \BibitemOpen
  \bibfield  {author} {\bibinfo {author} {\bibfnamefont {O.}~\bibnamefont
  {Benhar}}, \bibinfo {author} {\bibfnamefont {D.}~\bibnamefont {Day}},\ and\
  \bibinfo {author} {\bibfnamefont {I.}~\bibnamefont {Sick}},\ }\href
  {https://doi.org/10.1103/RevModPhys.80.189} {\bibfield  {journal} {\bibinfo
  {journal} {Rev. Mod. Phys.}\ }\textbf {\bibinfo {volume} {80}},\ \bibinfo
  {pages} {189} (\bibinfo {year} {2008})}\BibitemShut {NoStop}%
\bibitem [{\citenamefont {Nachtmann}(1997)}]{Nachtmann}%
  \BibitemOpen
  \bibfield  {author} {\bibinfo {author} {\bibfnamefont {O.}~\bibnamefont
  {Nachtmann}},\ }\href
  {https://doi.org/https://doi.org/10.1016/0550-3213(73)90144-2} {\bibfield
  {journal} {\bibinfo  {journal} {Nucl. Phys. B}\ }\textbf {\bibinfo {volume}
  {63}},\ \bibinfo {pages} {237} (\bibinfo {year} {1997})}\BibitemShut
  {NoStop}%
\bibitem [{\citenamefont {Jaffe}(1986)}]{Jaffe:1985je}%
  \BibitemOpen
  \bibfield  {author} {\bibinfo {author} {\bibfnamefont {R.~L.}\ \bibnamefont
  {Jaffe}},\ }in\ \href@noop {} {\emph {\bibinfo {booktitle} {{Relativistic
  Dynamics and Quark Nuclear Physics}}}},\ \bibinfo {editor} {edited by\
  \bibinfo {editor} {\bibfnamefont {M.~B.}\ \bibnamefont {Johnson}}\ and\
  \bibinfo {editor} {\bibfnamefont {A.}~\bibnamefont {Picklesimer}}}\ (\bibinfo
   {publisher} {Wiley-Interscience},\ \bibinfo {address} {New York},\ \bibinfo
  {year} {1986})\ \Eprint {https://arxiv.org/abs/2212.05616} {arXiv:2212.05616
  [hep-ph]} \BibitemShut {NoStop}%
\bibitem [{\citenamefont {Benhar}\ \emph {et~al.}(2006)\citenamefont {Benhar},
  \citenamefont {Day},\ and\ \citenamefont {Sick}}]{archive:qe}%
  \BibitemOpen
  \bibfield  {author} {\bibinfo {author} {\bibfnamefont {O.}~\bibnamefont
  {Benhar}}, \bibinfo {author} {\bibfnamefont {D.}~\bibnamefont {Day}},\ and\
  \bibinfo {author} {\bibfnamefont {I.}~\bibnamefont {Sick}},\ }\href
  {https://arxiv.org/abs/nucl-ex/0603032} {} (\bibinfo {year} {2006}),\ \Eprint
  {https://arxiv.org/abs/0603032} {arXiv:0603032 [nucl-ex]} \BibitemShut
  {NoStop}%
\bibitem [{\citenamefont {Benhar}\ \emph {et~al.}(1989)\citenamefont {Benhar},
  \citenamefont {Fabrocini},\ and\ \citenamefont {Fantoni}}]{Pke}%
  \BibitemOpen
  \bibfield  {author} {\bibinfo {author} {\bibfnamefont {O.}~\bibnamefont
  {Benhar}}, \bibinfo {author} {\bibfnamefont {A.}~\bibnamefont {Fabrocini}},\
  and\ \bibinfo {author} {\bibfnamefont {S.}~\bibnamefont {Fantoni}},\ }\href
  {https://doi.org/https://doi.org/10.1016/0375-9474(89)90374-6} {\bibfield
  {journal} {\bibinfo  {journal} {Nucl. Phys. A}\ }\textbf {\bibinfo {volume}
  {505}},\ \bibinfo {pages} {267} (\bibinfo {year} {1989})}\BibitemShut
  {NoStop}%
\bibitem [{\citenamefont {Benhar}\ \emph {et~al.}(1992)\citenamefont {Benhar},
  \citenamefont {Fabrocini},\ and\ \citenamefont {Fantoni}}]{Green}%
  \BibitemOpen
  \bibfield  {author} {\bibinfo {author} {\bibfnamefont {O.}~\bibnamefont
  {Benhar}}, \bibinfo {author} {\bibfnamefont {A.}~\bibnamefont {Fabrocini}},\
  and\ \bibinfo {author} {\bibfnamefont {S.}~\bibnamefont {Fantoni}},\ }\href
  {https://doi.org/https://doi.org/10.1016/0375-9474(92)90679-E} {\bibfield
  {journal} {\bibinfo  {journal} {Nucl. Phys. A}\ }\textbf {\bibinfo {volume}
  {550}},\ \bibinfo {pages} {201} (\bibinfo {year} {1992})}\BibitemShut
  {NoStop}%
\bibitem [{\citenamefont {Benhar}\ \emph {et~al.}(1994)\citenamefont {Benhar},
  \citenamefont {Fabrocini}, \citenamefont {Fantoni},\ and\ \citenamefont
  {Sick}}]{Benhar94}%
  \BibitemOpen
  \bibfield  {author} {\bibinfo {author} {\bibfnamefont {O.}~\bibnamefont
  {Benhar}}, \bibinfo {author} {\bibfnamefont {A.}~\bibnamefont {Fabrocini}},
  \bibinfo {author} {\bibfnamefont {S.}~\bibnamefont {Fantoni}},\ and\ \bibinfo
  {author} {\bibfnamefont {I.}~\bibnamefont {Sick}},\ }\href
  {https://doi.org/https://doi.org/10.1016/0375-9474(94)90920-2} {\bibfield
  {journal} {\bibinfo  {journal} {Nucl. Phys. A}\ }\textbf {\bibinfo {volume}
  {579}},\ \bibinfo {pages} {493} (\bibinfo {year} {1994})}\BibitemShut
  {NoStop}%
\bibitem [{\citenamefont {Ankowski}\ \emph {et~al.}(2024)\citenamefont
  {Ankowski}, \citenamefont {Benhar},\ and\ \citenamefont {Sakuda}}]{Ankowski}%
  \BibitemOpen
  \bibfield  {author} {\bibinfo {author} {\bibfnamefont {A.~M.}\ \bibnamefont
  {Ankowski}}, \bibinfo {author} {\bibfnamefont {O.}~\bibnamefont {Benhar}},\
  and\ \bibinfo {author} {\bibfnamefont {M.}~\bibnamefont {Sakuda}},\ }\href
  {https://doi.org/10.1103/PhysRevC.110.054612} {\bibfield  {journal} {\bibinfo
   {journal} {Phys. Rev. C}\ }\textbf {\bibinfo {volume} {110}},\ \bibinfo
  {pages} {054612} (\bibinfo {year} {2024})}\BibitemShut {NoStop}%
\bibitem [{\citenamefont {Koltun}(1972)}]{Koltun}%
  \BibitemOpen
  \bibfield  {author} {\bibinfo {author} {\bibfnamefont {D.~S.}\ \bibnamefont
  {Koltun}},\ }\href {https://doi.org/10.1103/PhysRevLett.28.182} {\bibfield
  {journal} {\bibinfo  {journal} {Phys. Rev. Lett.}\ }\textbf {\bibinfo
  {volume} {28}},\ \bibinfo {pages} {182} (\bibinfo {year} {1972})}\BibitemShut
  {NoStop}%
\bibitem [{\citenamefont {Harada}(1976)}]{Harada}%
  \BibitemOpen
  \bibfield  {author} {\bibinfo {author} {\bibfnamefont {M.}~\bibnamefont
  {Harada}},\ }\href {https://doi.org/https://doi.org/10.1007/BF02725916}
  {\bibfield  {journal} {\bibinfo  {journal} {Lett. Nuovo Cimento}\ }\textbf
  {\bibinfo {volume} {15}},\ \bibinfo {pages} {566} (\bibinfo {year}
  {1976})}\BibitemShut {NoStop}%
\bibitem [{\citenamefont {Carlson}\ \emph {et~al.}(2015)\citenamefont
  {Carlson}, \citenamefont {Gandolfi}, \citenamefont {Pederiva}, \citenamefont
  {Pieper}, \citenamefont {Schiavilla}, \citenamefont {Schmidt},\ and\
  \citenamefont {Wiringa}}]{Carlson:2014vla}%
  \BibitemOpen
  \bibfield  {author} {\bibinfo {author} {\bibfnamefont {J.}~\bibnamefont
  {Carlson}}, \bibinfo {author} {\bibfnamefont {S.}~\bibnamefont {Gandolfi}},
  \bibinfo {author} {\bibfnamefont {F.}~\bibnamefont {Pederiva}}, \bibinfo
  {author} {\bibfnamefont {S.~C.}\ \bibnamefont {Pieper}}, \bibinfo {author}
  {\bibfnamefont {R.}~\bibnamefont {Schiavilla}}, \bibinfo {author}
  {\bibfnamefont {K.~E.}\ \bibnamefont {Schmidt}},\ and\ \bibinfo {author}
  {\bibfnamefont {R.~B.}\ \bibnamefont {Wiringa}},\ }\href
  {https://doi.org/10.1103/RevModPhys.87.1067} {\bibfield  {journal} {\bibinfo
  {journal} {Rev. Mod. Phys.}\ }\textbf {\bibinfo {volume} {87}},\ \bibinfo
  {pages} {1067} (\bibinfo {year} {2015})}\BibitemShut {NoStop}%
\bibitem [{\citenamefont {Wiringa}\ \emph {et~al.}(1995)\citenamefont
  {Wiringa}, \citenamefont {Stoks},\ and\ \citenamefont
  {Schiavilla}}]{Wiringa:1994wb}%
  \BibitemOpen
  \bibfield  {author} {\bibinfo {author} {\bibfnamefont {R.~B.}\ \bibnamefont
  {Wiringa}}, \bibinfo {author} {\bibfnamefont {V.~G.~J.}\ \bibnamefont
  {Stoks}},\ and\ \bibinfo {author} {\bibfnamefont {R.}~\bibnamefont
  {Schiavilla}},\ }\href {https://doi.org/10.1103/PhysRevC.51.38} {\bibfield
  {journal} {\bibinfo  {journal} {Phys. Rev. C}\ }\textbf {\bibinfo {volume}
  {51}},\ \bibinfo {pages} {38} (\bibinfo {year} {1995})}\BibitemShut {NoStop}%
\bibitem [{\citenamefont {Pieper}(2008)}]{Pieper:2008rui}%
  \BibitemOpen
  \bibfield  {author} {\bibinfo {author} {\bibfnamefont {S.~C.}\ \bibnamefont
  {Pieper}},\ }\href {https://doi.org/10.1063/1.2932280} {\bibfield  {journal}
  {\bibinfo  {journal} {AIP Conf. Proc.}\ }\textbf {\bibinfo {volume} {1011}},\
  \bibinfo {pages} {143} (\bibinfo {year} {2008})}\BibitemShut {NoStop}%
\bibitem [{\citenamefont {Lonardoni}\ \emph {et~al.}(2017)\citenamefont
  {Lonardoni}, \citenamefont {Lovato}, \citenamefont {Pieper},\ and\
  \citenamefont {Wiringa}}]{Lonardoni:2017egu}%
  \BibitemOpen
  \bibfield  {author} {\bibinfo {author} {\bibfnamefont {D.}~\bibnamefont
  {Lonardoni}}, \bibinfo {author} {\bibfnamefont {A.}~\bibnamefont {Lovato}},
  \bibinfo {author} {\bibfnamefont {S.~C.}\ \bibnamefont {Pieper}},\ and\
  \bibinfo {author} {\bibfnamefont {R.~B.}\ \bibnamefont {Wiringa}},\ }\href
  {https://doi.org/10.1103/PhysRevC.96.024326} {\bibfield  {journal} {\bibinfo
  {journal} {Phys. Rev. C}\ }\textbf {\bibinfo {volume} {96}},\ \bibinfo
  {pages} {024326} (\bibinfo {year} {2017})}\BibitemShut {NoStop}%
\bibitem [{\citenamefont {Pudliner}\ \emph {et~al.}(1995)\citenamefont
  {Pudliner}, \citenamefont {Pandharipande}, \citenamefont {Carlson},\ and\
  \citenamefont {Wiringa}}]{Pudliner:1995wk}%
  \BibitemOpen
  \bibfield  {author} {\bibinfo {author} {\bibfnamefont {B.~S.}\ \bibnamefont
  {Pudliner}}, \bibinfo {author} {\bibfnamefont {V.~R.}\ \bibnamefont
  {Pandharipande}}, \bibinfo {author} {\bibfnamefont {J.}~\bibnamefont
  {Carlson}},\ and\ \bibinfo {author} {\bibfnamefont {R.~B.}\ \bibnamefont
  {Wiringa}},\ }\href {https://doi.org/10.1103/PhysRevLett.74.4396} {\bibfield
  {journal} {\bibinfo  {journal} {Phys. Rev. Lett.}\ }\textbf {\bibinfo
  {volume} {74}},\ \bibinfo {pages} {4396} (\bibinfo {year}
  {1995})}\BibitemShut {NoStop}%
\bibitem [{\citenamefont {Schmidt}\ and\ \citenamefont
  {Fantoni}(1999)}]{Schmidt:1999lik}%
  \BibitemOpen
  \bibfield  {author} {\bibinfo {author} {\bibfnamefont {K.~E.}\ \bibnamefont
  {Schmidt}}\ and\ \bibinfo {author} {\bibfnamefont {S.}~\bibnamefont
  {Fantoni}},\ }\href {https://doi.org/10.1016/S0370-2693(98)01522-6}
  {\bibfield  {journal} {\bibinfo  {journal} {Phys. Lett. B}\ }\textbf
  {\bibinfo {volume} {446}},\ \bibinfo {pages} {99} (\bibinfo {year}
  {1999})}\BibitemShut {NoStop}%
\bibitem [{\citenamefont {{I. Sick and D. Day}}(1992)}]{NM_EMC}%
  \BibitemOpen
  \bibfield  {author} {\bibinfo {author} {\bibnamefont {{I. Sick and D.
  Day}}},\ }\href
  {https://doi.org/https://doi.org/10.1016/0370-2693(92)90297-H} {\bibfield
  {journal} {\bibinfo  {journal} {Phys. Lett. B}\ }\textbf {\bibinfo {volume}
  {274}},\ \bibinfo {pages} {16} (\bibinfo {year} {1992})}\BibitemShut
  {NoStop}%
\bibitem [{\citenamefont {Day}\ \emph {et~al.}(1989)\citenamefont {Day},
  \citenamefont {McCarthy}, \citenamefont {Meziani}, \citenamefont {Minehart},
  \citenamefont {Sealock}, \citenamefont {Thornton}, \citenamefont {Jourdan},
  \citenamefont {Sick}, \citenamefont {Filippone}, \citenamefont {McKeown},
  \citenamefont {Milner}, \citenamefont {Potterveld},\ and\ \citenamefont
  {Szalata}}]{NM_response}%
  \BibitemOpen
  \bibfield  {author} {\bibinfo {author} {\bibfnamefont {D.~B.}\ \bibnamefont
  {Day}}, \bibinfo {author} {\bibfnamefont {J.~S.}\ \bibnamefont {McCarthy}},
  \bibinfo {author} {\bibfnamefont {Z.~E.}\ \bibnamefont {Meziani}}, \bibinfo
  {author} {\bibfnamefont {R.~C.}\ \bibnamefont {Minehart}}, \bibinfo {author}
  {\bibfnamefont {R.~M.}\ \bibnamefont {Sealock}}, \bibinfo {author}
  {\bibfnamefont {S.~T.}\ \bibnamefont {Thornton}}, \bibinfo {author}
  {\bibfnamefont {J.}~\bibnamefont {Jourdan}}, \bibinfo {author} {\bibfnamefont
  {I.}~\bibnamefont {Sick}}, \bibinfo {author} {\bibfnamefont {B.~W.}\
  \bibnamefont {Filippone}}, \bibinfo {author} {\bibfnamefont {R.~D.}\
  \bibnamefont {McKeown}}, \bibinfo {author} {\bibfnamefont {R.~G.}\
  \bibnamefont {Milner}}, \bibinfo {author} {\bibfnamefont {D.~H.}\
  \bibnamefont {Potterveld}},\ and\ \bibinfo {author} {\bibfnamefont
  {Z.}~\bibnamefont {Szalata}},\ }\href
  {https://doi.org/10.1103/PhysRevC.40.1011} {\bibfield  {journal} {\bibinfo
  {journal} {Phys. Rev. C}\ }\textbf {\bibinfo {volume} {40}},\ \bibinfo
  {pages} {1011} (\bibinfo {year} {1989})}\BibitemShut {NoStop}%
\bibitem [{\citenamefont {Akmal}\ and\ \citenamefont
  {Pandharipande}(1997)}]{AP}%
  \BibitemOpen
  \bibfield  {author} {\bibinfo {author} {\bibfnamefont {A.}~\bibnamefont
  {Akmal}}\ and\ \bibinfo {author} {\bibfnamefont {V.~R.}\ \bibnamefont
  {Pandharipande}},\ }\href {https://doi.org/10.1103/PhysRevC.56.2261}
  {\bibfield  {journal} {\bibinfo  {journal} {Phys. Rev. C}\ }\textbf {\bibinfo
  {volume} {56}},\ \bibinfo {pages} {2261} (\bibinfo {year}
  {1997})}\BibitemShut {NoStop}%
\bibitem [{\citenamefont {Marchand}\ \emph {et~al.}(1988)\citenamefont
  {Marchand}, \citenamefont {Bernheim}, \citenamefont {Dunn}, \citenamefont
  {G\'erard}, \citenamefont {Laget}, \citenamefont {Magnon}, \citenamefont
  {Morgenstern}, \citenamefont {Mougey}, \citenamefont {Picard}, \citenamefont
  {Reffay-Pikeroen}, \citenamefont {Turck-Chieze}, \citenamefont {Vernin},
  \citenamefont {Brussel}, \citenamefont {Capitani}, \citenamefont
  {De~Sanctis}, \citenamefont {Frullani},\ and\ \citenamefont
  {Garibaldi}}]{Marchand}%
  \BibitemOpen
  \bibfield  {author} {\bibinfo {author} {\bibfnamefont {C.}~\bibnamefont
  {Marchand}}, \bibinfo {author} {\bibfnamefont {M.}~\bibnamefont {Bernheim}},
  \bibinfo {author} {\bibfnamefont {P.~C.}\ \bibnamefont {Dunn}}, \bibinfo
  {author} {\bibfnamefont {A.}~\bibnamefont {G\'erard}}, \bibinfo {author}
  {\bibfnamefont {J.~M.}\ \bibnamefont {Laget}}, \bibinfo {author}
  {\bibfnamefont {A.}~\bibnamefont {Magnon}}, \bibinfo {author} {\bibfnamefont
  {J.}~\bibnamefont {Morgenstern}}, \bibinfo {author} {\bibfnamefont
  {J.}~\bibnamefont {Mougey}}, \bibinfo {author} {\bibfnamefont
  {J.}~\bibnamefont {Picard}}, \bibinfo {author} {\bibfnamefont
  {D.}~\bibnamefont {Reffay-Pikeroen}}, \bibinfo {author} {\bibfnamefont
  {S.}~\bibnamefont {Turck-Chieze}}, \bibinfo {author} {\bibfnamefont
  {P.}~\bibnamefont {Vernin}}, \bibinfo {author} {\bibfnamefont {M.~K.}\
  \bibnamefont {Brussel}}, \bibinfo {author} {\bibfnamefont {G.~P.}\
  \bibnamefont {Capitani}}, \bibinfo {author} {\bibfnamefont {E.}~\bibnamefont
  {De~Sanctis}}, \bibinfo {author} {\bibfnamefont {S.}~\bibnamefont
  {Frullani}},\ and\ \bibinfo {author} {\bibfnamefont {F.}~\bibnamefont
  {Garibaldi}},\ }\href {https://doi.org/10.1103/PhysRevLett.60.1703}
  {\bibfield  {journal} {\bibinfo  {journal} {Phys. Rev. Lett.}\ }\textbf
  {\bibinfo {volume} {60}},\ \bibinfo {pages} {1703} (\bibinfo {year}
  {1988})}\BibitemShut {NoStop}%
\bibitem [{\citenamefont {van Leeuwe}\ \emph {et~al.}(1998)\citenamefont {van
  Leeuwe}, \citenamefont {Blok}, \citenamefont {van~den Brand}, \citenamefont
  {Bulten}, \citenamefont {Dodge}, \citenamefont {Ent}, \citenamefont
  {Hesselink}, \citenamefont {Jans}, \citenamefont {Kasdorp}, \citenamefont
  {Laget}, \citenamefont {Lapik\'as}, \citenamefont {Nagorny}, \citenamefont
  {Onderwater}, \citenamefont {Pellegrino}, \citenamefont {Spaltro},
  \citenamefont {Steijger}, \citenamefont {Schiavilla}, \citenamefont
  {Templon},\ and\ \citenamefont {Unal}}]{Vanleeuwe}%
  \BibitemOpen
  \bibfield  {author} {\bibinfo {author} {\bibfnamefont {J.~J.}\ \bibnamefont
  {van Leeuwe}}, \bibinfo {author} {\bibfnamefont {H.~P.}\ \bibnamefont
  {Blok}}, \bibinfo {author} {\bibfnamefont {J.~F.~J.}\ \bibnamefont {van~den
  Brand}}, \bibinfo {author} {\bibfnamefont {H.~J.}\ \bibnamefont {Bulten}},
  \bibinfo {author} {\bibfnamefont {G.~E.}\ \bibnamefont {Dodge}}, \bibinfo
  {author} {\bibfnamefont {R.}~\bibnamefont {Ent}}, \bibinfo {author}
  {\bibfnamefont {W.~H.~A.}\ \bibnamefont {Hesselink}}, \bibinfo {author}
  {\bibfnamefont {E.}~\bibnamefont {Jans}}, \bibinfo {author} {\bibfnamefont
  {W.~J.}\ \bibnamefont {Kasdorp}}, \bibinfo {author} {\bibfnamefont {J.~M.}\
  \bibnamefont {Laget}}, \bibinfo {author} {\bibfnamefont {L.}~\bibnamefont
  {Lapik\'as}}, \bibinfo {author} {\bibfnamefont {S.~I.}\ \bibnamefont
  {Nagorny}}, \bibinfo {author} {\bibfnamefont {C.~J.~G.}\ \bibnamefont
  {Onderwater}}, \bibinfo {author} {\bibfnamefont {A.~R.}\ \bibnamefont
  {Pellegrino}}, \bibinfo {author} {\bibfnamefont {C.~M.}\ \bibnamefont
  {Spaltro}}, \bibinfo {author} {\bibfnamefont {J.~J.~M.}\ \bibnamefont
  {Steijger}}, \bibinfo {author} {\bibfnamefont {R.}~\bibnamefont
  {Schiavilla}}, \bibinfo {author} {\bibfnamefont {J.~A.}\ \bibnamefont
  {Templon}},\ and\ \bibinfo {author} {\bibfnamefont {O.}~\bibnamefont
  {Unal}},\ }\href {https://doi.org/10.1103/PhysRevLett.80.2543} {\bibfield
  {journal} {\bibinfo  {journal} {Phys. Rev. Lett.}\ }\textbf {\bibinfo
  {volume} {80}},\ \bibinfo {pages} {2543} (\bibinfo {year}
  {1998})}\BibitemShut {NoStop}%
\bibitem [{\citenamefont {Rohe}\ \emph {et~al.}(2004)\citenamefont {Rohe},
  \citenamefont {Armstrong}, \citenamefont {Asaturyan}, \citenamefont {Baker},
  \citenamefont {Bueltmann}, \citenamefont {Carasco}, \citenamefont {Day},
  \citenamefont {Ent}, \citenamefont {Fenker}, \citenamefont {Garrow},
  \citenamefont {Gasparian}, \citenamefont {Gueye}, \citenamefont {Hauger},
  \citenamefont {Honegger}, \citenamefont {Jourdan}, \citenamefont {Keppel},
  \citenamefont {Kubon}, \citenamefont {Lindgren}, \citenamefont {Lung},
  \citenamefont {Mack}, \citenamefont {Mitchell}, \citenamefont {Mkrtchyan},
  \citenamefont {Mocelj}, \citenamefont {Normand}, \citenamefont {Petitjean},
  \citenamefont {Rondon}, \citenamefont {Segbefia}, \citenamefont {Sick},
  \citenamefont {Stepanyan}, \citenamefont {Tang}, \citenamefont
  {Tiefenbacher}, \citenamefont {Vulcan}, \citenamefont {Warren}, \citenamefont
  {Wood}, \citenamefont {Yuan}, \citenamefont {Zeier}, \citenamefont {Zhu},\
  and\ \citenamefont {Zihlmann}}]{Rohe}%
  \BibitemOpen
  \bibfield  {author} {\bibinfo {author} {\bibfnamefont {D.}~\bibnamefont
  {Rohe}}, \bibinfo {author} {\bibfnamefont {C.~S.}\ \bibnamefont {Armstrong}},
  \bibinfo {author} {\bibfnamefont {R.}~\bibnamefont {Asaturyan}}, \bibinfo
  {author} {\bibfnamefont {O.~K.}\ \bibnamefont {Baker}}, \bibinfo {author}
  {\bibfnamefont {S.}~\bibnamefont {Bueltmann}}, \bibinfo {author}
  {\bibfnamefont {C.}~\bibnamefont {Carasco}}, \bibinfo {author} {\bibfnamefont
  {D.}~\bibnamefont {Day}}, \bibinfo {author} {\bibfnamefont {R.}~\bibnamefont
  {Ent}}, \bibinfo {author} {\bibfnamefont {H.~C.}\ \bibnamefont {Fenker}},
  \bibinfo {author} {\bibfnamefont {K.}~\bibnamefont {Garrow}}, \bibinfo
  {author} {\bibfnamefont {A.}~\bibnamefont {Gasparian}}, \bibinfo {author}
  {\bibfnamefont {P.}~\bibnamefont {Gueye}}, \bibinfo {author} {\bibfnamefont
  {M.}~\bibnamefont {Hauger}}, \bibinfo {author} {\bibfnamefont
  {A.}~\bibnamefont {Honegger}}, \bibinfo {author} {\bibfnamefont
  {J.}~\bibnamefont {Jourdan}}, \bibinfo {author} {\bibfnamefont {C.~E.}\
  \bibnamefont {Keppel}}, \bibinfo {author} {\bibfnamefont {G.}~\bibnamefont
  {Kubon}}, \bibinfo {author} {\bibfnamefont {R.}~\bibnamefont {Lindgren}},
  \bibinfo {author} {\bibfnamefont {A.}~\bibnamefont {Lung}}, \bibinfo {author}
  {\bibfnamefont {D.~J.}\ \bibnamefont {Mack}}, \bibinfo {author}
  {\bibfnamefont {J.~H.}\ \bibnamefont {Mitchell}}, \bibinfo {author}
  {\bibfnamefont {H.}~\bibnamefont {Mkrtchyan}}, \bibinfo {author}
  {\bibfnamefont {D.}~\bibnamefont {Mocelj}}, \bibinfo {author} {\bibfnamefont
  {K.}~\bibnamefont {Normand}}, \bibinfo {author} {\bibfnamefont
  {T.}~\bibnamefont {Petitjean}}, \bibinfo {author} {\bibfnamefont
  {O.}~\bibnamefont {Rondon}}, \bibinfo {author} {\bibfnamefont
  {E.}~\bibnamefont {Segbefia}}, \bibinfo {author} {\bibfnamefont
  {I.}~\bibnamefont {Sick}}, \bibinfo {author} {\bibfnamefont {S.}~\bibnamefont
  {Stepanyan}}, \bibinfo {author} {\bibfnamefont {L.}~\bibnamefont {Tang}},
  \bibinfo {author} {\bibfnamefont {F.}~\bibnamefont {Tiefenbacher}}, \bibinfo
  {author} {\bibfnamefont {W.~F.}\ \bibnamefont {Vulcan}}, \bibinfo {author}
  {\bibfnamefont {G.}~\bibnamefont {Warren}}, \bibinfo {author} {\bibfnamefont
  {S.~A.}\ \bibnamefont {Wood}}, \bibinfo {author} {\bibfnamefont
  {L.}~\bibnamefont {Yuan}}, \bibinfo {author} {\bibfnamefont {M.}~\bibnamefont
  {Zeier}}, \bibinfo {author} {\bibfnamefont {H.}~\bibnamefont {Zhu}},\ and\
  \bibinfo {author} {\bibfnamefont {B.}~\bibnamefont {Zihlmann}} (\bibinfo
  {collaboration} {E97-006 Collaboration}),\ }\href
  {https://doi.org/10.1103/PhysRevLett.93.182501} {\bibfield  {journal}
  {\bibinfo  {journal} {Phys. Rev. Lett.}\ }\textbf {\bibinfo {volume} {93}},\
  \bibinfo {pages} {182501} (\bibinfo {year} {2004})}\BibitemShut {NoStop}%
\bibitem [{\citenamefont {Gnech}\ \emph {et~al.}(2025)\citenamefont {Gnech},
  \citenamefont {Lovato},\ and\ \citenamefont {Rocco}}]{Gnech:2024qru}%
  \BibitemOpen
  \bibfield  {author} {\bibinfo {author} {\bibfnamefont {A.}~\bibnamefont
  {Gnech}}, \bibinfo {author} {\bibfnamefont {A.}~\bibnamefont {Lovato}},\ and\
  \bibinfo {author} {\bibfnamefont {N.}~\bibnamefont {Rocco}},\ }\href
  {https://doi.org/10.1103/PhysRevC.111.024314} {\bibfield  {journal} {\bibinfo
   {journal} {Phys. Rev. C}\ }\textbf {\bibinfo {volume} {111}},\ \bibinfo
  {pages} {024314} (\bibinfo {year} {2025})}\BibitemShut {NoStop}%
\bibitem [{\citenamefont {Wiringa}\ and\ \citenamefont {Pieper}(2002)}]{V8P}%
  \BibitemOpen
  \bibfield  {author} {\bibinfo {author} {\bibfnamefont {R.}~\bibnamefont
  {Wiringa}}\ and\ \bibinfo {author} {\bibfnamefont {S.}~\bibnamefont
  {Pieper}},\ }\href {https://doi.org/10.1103/PhysRevLett.89.182501} {\bibfield
   {journal} {\bibinfo  {journal} {Phys. Rev. Lett.}\ }\textbf {\bibinfo
  {volume} {89}},\ \bibinfo {pages} {182501} (\bibinfo {year}
  {2002})}\BibitemShut {NoStop}%
\bibitem [{\citenamefont {Lovato}\ \emph {et~al.}(2022)\citenamefont {Lovato},
  \citenamefont {Bombaci}, \citenamefont {Logoteta}, \citenamefont {Piarulli},\
  and\ \citenamefont {Wiringa}}]{AFDMC_PNM}%
  \BibitemOpen
  \bibfield  {author} {\bibinfo {author} {\bibfnamefont {A.}~\bibnamefont
  {Lovato}}, \bibinfo {author} {\bibfnamefont {I.}~\bibnamefont {Bombaci}},
  \bibinfo {author} {\bibfnamefont {D.}~\bibnamefont {Logoteta}}, \bibinfo
  {author} {\bibfnamefont {M.}~\bibnamefont {Piarulli}},\ and\ \bibinfo
  {author} {\bibfnamefont {R.~B.}\ \bibnamefont {Wiringa}},\ }\href
  {https://doi.org/10.1103/PhysRevC.105.055808} {\bibfield  {journal} {\bibinfo
   {journal} {Phys. Rev. C}\ }\textbf {\bibinfo {volume} {105}},\ \bibinfo
  {pages} {055808} (\bibinfo {year} {2022})}\BibitemShut {NoStop}%
\bibitem [{\citenamefont {Piarulli}\ \emph {et~al.}(2018)\citenamefont
  {Piarulli}, \citenamefont {Baroni}, \citenamefont {Girlanda}, \citenamefont
  {Kievsky}, \citenamefont {Lovato}, \citenamefont {Lusk}, \citenamefont
  {Marcucci}, \citenamefont {Pieper}, \citenamefont {Schiavilla}, \citenamefont
  {Viviani},\ and\ \citenamefont {Wiringa}}]{Piarulli:2017dwd}%
  \BibitemOpen
  \bibfield  {author} {\bibinfo {author} {\bibfnamefont {M.}~\bibnamefont
  {Piarulli}}, \bibinfo {author} {\bibfnamefont {A.}~\bibnamefont {Baroni}},
  \bibinfo {author} {\bibfnamefont {L.}~\bibnamefont {Girlanda}}, \bibinfo
  {author} {\bibfnamefont {A.}~\bibnamefont {Kievsky}}, \bibinfo {author}
  {\bibfnamefont {A.}~\bibnamefont {Lovato}}, \bibinfo {author} {\bibfnamefont
  {E.}~\bibnamefont {Lusk}}, \bibinfo {author} {\bibfnamefont {L.~E.}\
  \bibnamefont {Marcucci}}, \bibinfo {author} {\bibfnamefont {S.~C.}\
  \bibnamefont {Pieper}}, \bibinfo {author} {\bibfnamefont {R.}~\bibnamefont
  {Schiavilla}}, \bibinfo {author} {\bibfnamefont {M.}~\bibnamefont
  {Viviani}},\ and\ \bibinfo {author} {\bibfnamefont {R.~B.}\ \bibnamefont
  {Wiringa}},\ }\href {https://doi.org/10.1103/PhysRevLett.120.052503}
  {\bibfield  {journal} {\bibinfo  {journal} {Phys. Rev. Lett.}\ }\textbf
  {\bibinfo {volume} {120}},\ \bibinfo {pages} {052503} (\bibinfo {year}
  {2018})}\BibitemShut {NoStop}%
\bibitem [{\citenamefont {Lynn}\ \emph {et~al.}(2016)\citenamefont {Lynn},
  \citenamefont {Tews}, \citenamefont {Carlson}, \citenamefont {Gandolfi},
  \citenamefont {Gezerlis}, \citenamefont {Schmidt},\ and\ \citenamefont
  {Schwenk}}]{Lynn:2015jua}%
  \BibitemOpen
  \bibfield  {author} {\bibinfo {author} {\bibfnamefont {J.~E.}\ \bibnamefont
  {Lynn}}, \bibinfo {author} {\bibfnamefont {I.}~\bibnamefont {Tews}}, \bibinfo
  {author} {\bibfnamefont {J.}~\bibnamefont {Carlson}}, \bibinfo {author}
  {\bibfnamefont {S.}~\bibnamefont {Gandolfi}}, \bibinfo {author}
  {\bibfnamefont {A.}~\bibnamefont {Gezerlis}}, \bibinfo {author}
  {\bibfnamefont {K.~E.}\ \bibnamefont {Schmidt}},\ and\ \bibinfo {author}
  {\bibfnamefont {A.}~\bibnamefont {Schwenk}},\ }\href
  {https://doi.org/10.1103/PhysRevLett.116.062501} {\bibfield  {journal}
  {\bibinfo  {journal} {Phys. Rev. Lett.}\ }\textbf {\bibinfo {volume} {116}},\
  \bibinfo {pages} {062501} (\bibinfo {year} {2016})}\BibitemShut {NoStop}%
\end{thebibliography}
%


\section*{End Matter}

%

In this Section, we briefly discuss the accuracy of the variational estimates of the nuclear ground-state energies. In addition, in order to illustrate the dependence of 
our analysis on the model of nuclear dynamics, we report the results obtained using an Hamiltonian model derived from $\chi$EFT.~\\
 
\paragraph*{Accuracy of CVMC and FHNC results}
To assess the reliability of the CVMC results\textemdash based on a cluster expansion of the expectation values of the 
kinetic and potential-energy operators\textemdash we have compared the value of $\langle E_A \rangle$ computed from Eq.~\eqref{Koltun} using 
the variational estimate of the ground-state energy of ${\isotope[16][]{O}}$ to the result of a highly-accurate 
calculation performed within the AFDMC scheme~\cite{Schmidt:1999lik}. The CVMC value, $\approx 44.1$~MeV, turns out to be 
in excellent agreement with the corresponding AFDMC result, $\approx43.7$ MeV, obtained by the authors of Ref.~\cite{Gnech:2024qru} using the Argonne $v_{8}^\prime$ plus Urbana IX Hamiltonian. The small discrepancy can be attributed, at least in part, to differences between the  full Argonne~$v_{18}$ potential and the re-projected Argonne $v_{8}^\prime$ model~\cite{V8P}.

In the case of SNM, AFDMC results obtained from the $v_{18}$ plus Urbana IX Hamiltonian employed in
Ref.~\cite{AP}  are not available. However, the accuracy of the variational approach is 
strongly supported by the study of pure neutron matter (PNM) carried out by the authors of Ref.~\cite{AFDMC_PNM}. At nuclear matter equilibrium density, corresponding to 0.16 fm$^{-3}$, the ground-state energy 
of PNM reported in this paper turns out be in perfect agreement with the FHNC result of Ref.~\cite{AP}.~\\ 

\paragraph*{Data analysis with $\chi$EFT removal energies}
In the main text, we argued that the linear dependence of the EMC effect on $\langle E_A \rangle$ does not depend significantly on the specific model
of nuclear dynamics. To strengthen this statement, we have performed calculations of the average removal energies using two sets of $\chi$EFT interactions that are local in coordinate space. 

In Table~\ref{tab:koltun_summary_chiralEFT}, we present GFMC results for \isotope[2][]{H}, \isotope[3][]{He}, \isotope[4][]{He}, \isotope[8][]{Be}, and 
\isotope[12][]{C}, obtained from the $\Delta$-full $\chi$EFT Hamiltonian of Ref.~\cite{Piarulli:2017dwd}. Specifically, we have employed the NV2+3-Ia interaction, which reproduces the spectra of nuclei up to  \isotope[12][]{C} with great accuracy. The deviations from the removal energies of nuclei with 
$A \leq 12$ listed in Table~\ref{tab:koltun_summary} are due to the lower expectation values of the kinetic energies, which in turn 
reflect the comparatively weaker correlation effects predicted by dynamical models based on $\chi$EFT.

\begin{table}[!t]
\centering
\caption{\small Same as  in Table~\ref{tab:koltun_summary}, but for the local $\Delta$-full $\chi$EFT Hamiltonian NV2+3-Ia~~\cite{Piarulli:2017dwd}. 
All calculations have been performed with the GFMC method. Energies are in MeV.
}
\label{tab:koltun_summary_chiralEFT}
\vspace*{.10in}
\begin{tabular}{lccccc}
\hline\hline
Nucleus & \ \ ${\mathcal E}_A/A$ & \ \ $\langle T\rangle/A$ & \ \ \ $\langle V_{3N}\rangle/A$ & \ \ $\langle E_A \rangle$ \\
\hline
{\isotope[2][]{H}}   & $-1.113$ & $9.905$ & $0.000$  & $2.225$ \\
{\isotope[3][]{He}}  & $-2.571$ & $12.950$ & $-0.305$ & $11.921$ \\
{\isotope[4][]{He}}  & $-7.050$ & $21.029$ & $-1.280$ & $29.399$ \\
{\isotope[9][]{Be}}  & $-6.562$ & $24.742$ & $-2.066$ & $36.841$ \\
{\isotope[10][]{B}}  & $-6.539$ & $26.178$ & $-2.581$ & $38.928$ \\
{\isotope[11][]{B}}  & $-7.123$ & $25.491$ & $-2.373$ & $39.561$ \\
{\isotope[12][]{C}}  & $-7.848$ & $27.536$ & $-2.664$ & $43.392$ \\
\hline
\end{tabular}
\end{table}

\begin{figure}[!htb]

\includegraphics[scale=0.65]{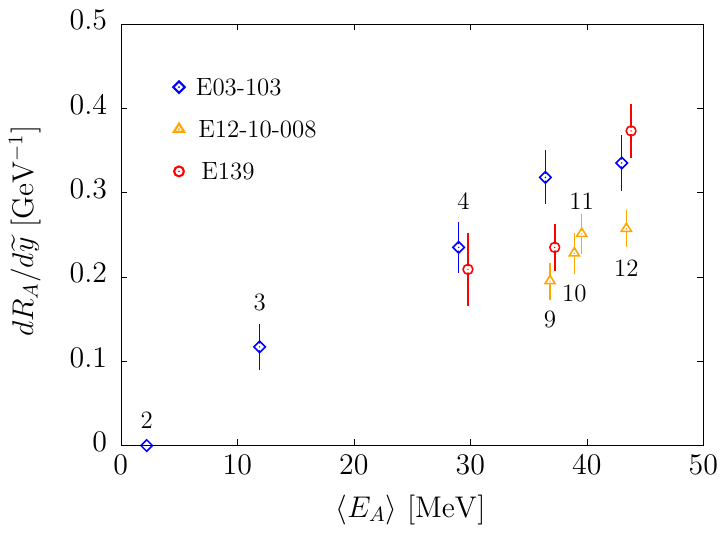}
\vspace*{-0.15in}
\caption{\small Slopes of the EMC ratios in the ${\widetilde y}$ region corresponding to $0.35 < x < 0.7$, 
labelled according to the nuclear mass number $A$. The average removal energies are obtained with the $\Delta$-full $\chi$EFT Hamiltonian NV2+3-Ia. For clarity, the E139 points and the E03-103 points corresponding to $A=$ 4, 9 and 12 are offset by $\pm 0.4$ MeV, respectively.
\label{EMC:shift_chiral}}
\end{figure}

In Fig.~\ref{EMC:shift_chiral}, the slope of the EMC ratios in the $\widetilde y$ region corresponding to $0.35 < x < 0.7$ is displayed 
as a function of the average removal energies reported in Table~\ref{tab:koltun_summary_chiralEFT}, obtained with the $\Delta$-full $\chi$EFT Hamiltonian NV2+3-Ia. The linear correlation is still clearly observed, although, compared to Fig.~\ref{tab:koltun_summary}, the different set of $\langle E_A \rangle$ values appears to bring about a somewhat larger slope.  

For the sake of completeness, Table~\ref{tab:koltun_summary_chiralEFT_2} reports the results of AFDMC calculations of \isotope[3][]{He}, \isotope[4][]{He} and \isotope[12][]{C} energies, 
performed with the local $\chi$EFT interaction of Ref.~\cite{Lynn:2015jua}, obtained setting the cutoff  to $R_0 = 1.0$~fm and  retaining only the 
component of the three-nucleon-force labelled $E_\tau$. As in the case of CVMC, for all nuclei 
we also provide results obtained by replacing the AFDMC ground-state energies with the corresponding experimental values, while keeping the kinetic-energy and three-body-force expectation values unchanged.

\begin{table}[H]
\centering
\caption{\small
Same as in Tables~\ref{tab:koltun_summary} and~\ref{tab:koltun_summary_chiralEFT}, but for the local $\chi$EFT interaction of Ref.~\cite{Lynn:2015jua}. The cutoff is set to is $R_0 = 1.0$ fm, and the three-body potential only includes the contribution labelled $E_\tau$. All calculations have been carried out with the AFDMC method and the resulting energies are given in MeV. 
We also report results obtained from Eq.~\eqref{Koltun} by replacing the AFDMC ground-state energies with the corresponding experimental values, while keeping the same kinetic and $V_{3N}$ contributions.}
\label{tab:koltun_summary_chiralEFT_2}
\vspace*{.10in}
\begin{tabular}{lcccc}
\hline\hline
Nucleus & \ \ ${\mathcal E}_A/A$ & \ \ $\langle T\rangle/A$ & \ \ \ $\langle V_{3N}\rangle/A$ & \ \ $\langle E_A \rangle$ \\
\hline
\multirow{2}{*}{$^{3}$He}   & $-2.517$ & $15.747$ & $-0.340$ & $13.247$ \\
           & $-2.573$ & $15.747$ & $-0.340$ & $13.359$ \\
\hline
\multirow{2}{*}{$^{4}$He}   & $-6.910$ & $24.545$ & $-1.960$ & $32.143$ \\
& $-7.074$ & $24.545$ & $-1.960$ & $32.471$ \\
\hline
\multirow{2}{*}{$^{12}$C}   & $-6.500$ & $25.667$ & $-2.872$ & $39.205$ \\
& $-7.680$ & $25.667$ & $-2.872$ & $41.565$ \\
\hline\hline
\end{tabular}
\end{table}

%


\end{document}